\documentclass[12pt, a4paper]{article}
\pdfoutput=1
\usepackage{amsmath}
\usepackage{amsfonts}
\usepackage{amssymb}
\usepackage{bbm}
\usepackage{verbatim}
\usepackage{dsfont}
\usepackage{booktabs}
\usepackage{slashed}

\usepackage{epsfig}
\usepackage{color}
\usepackage[table]{xcolor}
\usepackage{graphicx}
\usepackage[footnotesize]{caption}
\usepackage[listofformat=empty,subrefformat=empty]{subfig}

\usepackage{float}
\usepackage{overpic}

\usepackage{enumerate}
\usepackage{hhline}
\usepackage{multirow}

\usepackage{cite}
\usepackage{xspace}
\usepackage{setspace}

\newcommand{\be}{\begin{equation}}
\newcommand{\ee}{\end{equation}}
\newcommand{\ol}[1]{\overline{#1}}
\newcommand{\hc}{+\,\mathrm{h.c.}}

\newcommand{\SU}[1]{\ensuremath{\mathrm{SU}(#1)}}
\newcommand{\U}[1]{\ensuremath{\mathrm{U}(#1)}}
\newcommand{\into}{\ensuremath{\,\rightarrow\,}}

\newcommand{\MSbar}{\ensuremath{\ol{\text{MS}}}}
\newcommand{\DRbar}{\ensuremath{\ol{\text{DR}}}}
\newcommand{\SM}{\text{SM}}
\newcommand{\THDM}{\text{THDM}}
\newcommand{\unit}[1]{\;\text{#1}}
\DeclareMathOperator{\re}{\mathrm{Re}}
\DeclareMathOperator{\im}{\mathrm{Im}}
\newcommand{\lsim}
{\;\raisebox{-.3em}{$\stackrel{\displaystyle <}{\sim}$}\;}

\begin{document}

\thispagestyle{empty}

\begin{flushleft}
DESY-15-238
\end{flushleft}
\vskip 2 cm
\begin{center}
{\Large {\bf Vacuum stability and supersymmetry at high scales\\[.3em]
with two Higgs doublets}}\\[10pt]
\bigskip
\bigskip
{
E.~Bagnaschi$^a$, F.~Br\"ummer$^b$, W.~Buchm\"uller$^a$, A.~Voigt$^a$,
G.~Weiglein$^a$
}
\bigskip
\vspace{0.23cm}\\
{\it\small $^a$Deutsches Elektronen-Synchrotron DESY, 22607 Hamburg, Germany\\
$^b$Laboratoire Univers et Particules de Montpellier, UMR5299,\\
Universit\'e de Montpellier, 34095 Montpellier, France}\\[20pt]
\bigskip
\end{center}

\begin{abstract}
\noindent
We investigate the stability of the electroweak vacuum for two-Higgs-doublet models
with a supersymmetric UV completion.
The supersymmetry breaking scale is taken to be of the order of the
grand unification scale. We first study the case where all
superpartners decouple at this scale.
We show that contrary to the Standard Model with one Higgs doublet,
matching to the supersymmetric UV completion is possible if the
low-scale model contains two Higgs doublets.
In this case vacuum stability and experimental constraints
point towards low values of $\tan\beta\lesssim 2$ and pseudoscalar
masses of at least about a TeV. If the higgsino superpartners of the
Higgs fields are also kept light, the conclusions are similar and
essentially independent of the higgsino mass.
Finally, if all gauginos are also given electroweak-scale masses
(split supersymmetry with two Higgs doublets), the model cannot be
matched to supersymmetry at very high scales when requiring a 125 GeV
Higgs. Light neutral and charged higgsinos therefore emerge as a
promising signature of a supersymmetric UV completion of the
Standard Model at the grand unification scale.
\end{abstract}

\newpage
\setcounter{page}{2}
\setcounter{footnote}{0}

\onehalfspacing

\section{Introduction}
\label{sec: Introduction}

The structure of the electroweak and strong interactions seems to
point towards
an increase of symmetry and to a unification of the fundamental forces
as we probe shorter and shorter distances. It is then natural to
expect that symmetries larger than the internal and space-time
symmetries of the Standard Model of particle physics, including supersymmetry,
grand unification and additional space-time dimensions, will play a crucial
role for the embedding of the Standard Model into a more fundamental theory.
In particular string theory, the leading candidate for a unified theory of all
interactions, relies on supersymmetry to guarantee a perturbatively controlled
stable vacuum state \cite{Green:1987mn,Ibanez:2012zz}.
From the point of view of superstring theory, the generic expectation for the
scale of supersymmetry breaking is at or close to the string scale, which is of
course usually many orders of magnitude larger than the electroweak scale. The
Standard Model, possibly supplemented by other light states, would then be the
non-supersymmetric effective field theory of a UV completion with spontaneously
broken supersymmetry. This UV completion would take effect at a very high energy
scale of about $10^{15-17}$ GeV.
Example scenarios include universal high-scale supersymmetry \cite{Hall:2009nd} and split
supersymmetry \cite{ArkaniHamed:2004fb,Giudice:2004tc}, which has been
realised in string theory \cite{Antoniadis:2004dt} and
higher-dimensional field theory with flux \cite{Buchmuller:2015jna}.

In the past, the main motivation to consider supersymmetric extensions
of the Standard Model used to be the hierarchy problem:
electroweak-scale supersymmetry allows to stabilise a large
hierarchy between the electroweak scale and a much higher fundamental scale
against radiative corrections.
However, so far the data shows no sign of supersymmetry.
Should no evidence in its favour surface during the second run of the LHC,
one may have to conclude
that the electroweak scale is not actually protected by supersymmetry, but fixed by some
unknown ultraviolet dynamics at a value which presently appears unnatural to us.
Our hypothesis for the present paper is that supersymmetry does exist but, since the scale of
its breaking is high, that it plays no role in stabilising the electroweak hierarchy.

Admitting a supersymmetric UV completion at high scales is a nontrivial
constraint on the low-energy effective theory. For example, it is well known
that the Standard Model by itself cannot be matched to its minimal
supersymmetric extension (MSSM)
above about $10^{11}$ GeV \cite{Giudice:2011cg}. This
is because at higher energies the running Higgs quartic coupling in the Standard
Model becomes negative, while the D-term potential in supersymmetry is positive
definite. The maximal matching scale is even lower for split supersymmetry,
where the electroweak-scale spectrum consists of the Standard Model and the MSSM
gauginos and higgsinos \cite{Giudice:2011cg, Bagnaschi:2014rsa}. Therefore, to
allow for a supersymmetric UV completion at scales of $10^{15-17}$ GeV, more
states need to be kept light in the low-energy theory, besides the Standard
Model Higgs doublet and possibly gauginos and higgsinos.

Our ability to extrapolate some non-supersymmetric low-energy effective theory
to high energies may also be limited by vacuum stability. This, again, is
already seen in the Standard Model itself: as a result of the quartic coupling
turning negative, the Higgs potential becomes unbounded from below (although
the lifetime of the electroweak vacuum has been estimated to be longer than the
age of the universe, see Ref.~\cite{Buttazzo:2013uya} and references therein). 
More generally, demanding a stable
or at least sufficiently long-lived vacuum imposes additional constraints on any
low-energy theory, even if it can be matched to a supersymmetric UV completion.
Although supersymmetry ensures that the potential will be positive definite at
the UV completion scale, the RG-improved tree-level potential may still be
formally unbounded from below at intermediate energies when expressed in terms
of the running couplings. This would signal the presence of additional vacua
which are in general deeper than the realistic electroweak vacuum.

A particularly interesting class of models retains both MSSM Higgs doublets as
light states at low energies, with or without the light higgsinos and gauginos
of split supersymmetry. The matching of the two-Higgs-doublet model (THDM) to
 the MSSM at high energies has previously been discussed in
Ref.~\cite{Gorbahn:2009pp}. Recently, a detailed
analysis of the matching for a variety of THDM models
as function of the supersymmetry breaking scale has been performed in
Ref.~\cite{Lee:2015uza}, however, without taking vacuum stability constraints
into account. With regards to vacuum stability, the extrapolation of a THDM to high energies near the
Planck scale was studied in
Refs~\cite{Chakrabarty:2014aya,Das:2015mwa, Chowdhury:2015yja,
  Ferreira:2015rha},
but without imposing constraints from high-energy supersymmetry.

In the present paper we show that several kinds of two-Higgs-doublet models can
indeed be matched to GUT-scale or even to  string-scale supersymmetry without
suffering from vacuum instability. We study three exemplary models
using the spectrum generator framework FlexibleSUSY~\cite{Athron:2014yba}: a pure
type-II THDM, the THDM with additional electroweak-scale higgsinos (which has
the appealing property of gauge coupling unification at $10^{14}$ GeV), and the
THDM with the full gaugino and higgsino field content of split supersymmetry at the
electroweak scale. It turns out that the combined requirements of a
supersymmetric UV completion, a stable vacuum, and a 125 GeV Higgs are quite
restrictive on the low-energy spectrum. For the pure THDM we find that
the parameter region at low $\tan\beta$ and relatively large $M_A$,
namely $\tan\beta\lesssim 2$ and $M_A\gtrsim 1$~TeV, is in agreement with
all these constraints as well as with the experimental bounds 
from the measurement of ${\rm BR}(b\into s\gamma)$~\cite{Amhis:2014hma}
and the limits from
the searches for additional Higgs bosons, in particular in the channel
$H,A\into\tau\tau$~\cite{Aad:2014vgg,CMS:2015mca,ATLASHtautau13TeV}.
The conclusions are similar but somewhat more restrictive for the THDM with
light Higgsinos. For the THDM with split supersymmetry, on the other
hand, we find that the model
cannot be extrapolated to the scale of Grand Unification because the predicted
mass of the Standard Model-like Higgs boson is always too large in the
parameter regions allowed by the other constraints.

\section{THDM models as effective field theories}

\subsection{Preliminary remarks}

The standard procedure for treating theories with several hierarchically
separated scales is to ``run and match'' the effective field theory
parameters. That is, the theory is regularised and renormalised using the
$\ol{\rm MS}$ scheme (or one of its cousins such as
$\ol{\rm DR}$), the parameters are evolved according to their $n$-loop
renormalisation group equations in between the thresholds, and at each
threshold crossing the heavy states are decoupled by hand. The parameters
of the resulting effective theory are matched to those of the full theory
with $(n-1)$-loop precision. If the masses of two heavy states are
comparable to each other,
they should be decoupled simultaneously and their mass difference accounted
for by an appropriate threshold correction at leading-log order. If on the
other hand the masses of two heavy states are widely separated, then they
define two distinct thresholds between which the logarithms should be
resummed, using the renormalisation group equations of an intermediate
effective theory.

For the present study we will always use precisely one effective field
theory between the supersymmetry breaking scale $M_S=10^{15-17}$ GeV and
the electroweak scale. While intermediate thresholds certainly offer
interesting possibilities to generalise our work, here we will always
assume that one set of particles decouples close to $M_S$ and that the
remaining states will obtain masses at most of the order of a TeV. These
``light'' states will always include the Standard Model particles and
a second Higgs doublet; we will furthermore
investigate the cases where they also
include a pair of higgsinos, or a pair of higgsinos and all MSSM
gauginos.

In particular, we take all the eigenvalues of the Higgs mass matrix to be
comparable to each other,
and therefore the running parameters of the THDM must be matched
directly to the measured pole masses of the Standard Model particles.
Thus, our study differs from the often considered case where the mass
scale $M_A$ of the non-standard Higgs bosons is much higher than the 
electroweak scale. In this case
the appropriate procedure would be to decouple the
non-standard Higgs bosons at the high scale $M_A$,
to match the THDM to the Standard Model at $M_A$, and then to evolve
the Standard Model running parameters down to the electroweak scale.

Imposing that all Higgs bosons acquire masses $\lesssim$ TeV is a strong
assumption, which as discussed above is technically unnatural since as
for the discovered Higgs boson at 125~GeV also the masses of further
relatively light Higgs bosons should be affected by high-scale physics.
The Higgs mass parameters of the
low-energy theory are determined by the matching conditions to the unknown
supersymmetric theory at $M_S\sim 10^{15-17}$, and are generically
expected to be of the order
of $M_S$ itself. Here we postulate that the various contributions to the Higgs
mass matrix cancel each other to a very high degree of precision, such that all of 
its entries are of the order of at most a TeV. We refrain from speculating about 
the reasons --- in our approach we assume that the hierarchy problem is solved 
by the UV theory by some means unknown to us. It has been argued that the
electroweak scale might need to be low for anthropic reasons, and that this
would predict precisely one light scalar doublet. We do not subscribe to these
arguments; it seems to us that they rest on rather frail assumptions, and that
even if anthropics should indeed be related to
the electroweak hierarchy, this would not necessarily
preclude a (presently unknown) anthropic argument for a second light Higgs
doublet.

\subsection{Conventions for the THDM}

We use the following conventions for
parameterising the scalar potential of the
THDM as
\be
\begin{split}
V=&\;
m_1^2\,H_1^\dag H_1+m_2^2\,H_2^\dag H_2-\left(m_{12}^2\,H_1^\dag
H_2\hc\right)+V_4\,,\\
V_4=&\;\frac{\lambda_1}{2}(H^\dag_1 H_1)^2
 +\frac{\lambda_2}{2}(H^\dag_2 H_2)^2+\lambda_3(H^\dag_1 H_1)(H^\dag_2 H_2)
 +\lambda_4|H_1^\dag H_2|^2\\
&+\left(\frac{\lambda_5}{2} (H_1^\dag H_2)^2
 + \lambda_6 (H_1^\dag H_2)(H_1^\dag H_1)
 +\lambda_7(H_1^\dag H_2)(H_2^\dag H_2)    \hc\right)\,.
\end{split}
\ee
For each Yukawa term allowed in the Standard Model, the general THDM contains
two such terms, one involving $H_1$ and the other involving $H_2$.
Moreover, if there are light gauginos ($\tilde{B}$, $\tilde{W}^i$,
$\tilde{G}^a$) and higgsinos ($\tilde{h}_d$, $\tilde{h}_u$) in the
spectrum, they are coupled to the Higgs doublets with the Yukawa terms
\be
\begin{split}
-{\cal L}_{\rm Yuk} = &\;
 \frac{\tilde g_{d}}{\sqrt{2}} H_1\tilde W\tilde h_d
 +\frac{\tilde g_{d}'}{\sqrt{2}} H_1\tilde B\tilde h_d
 +\frac{\tilde g_{u}}{\sqrt{2}}H_2^\dag\tilde W\tilde h_u
 +\frac{\tilde g_{u}'}{\sqrt{2}}H_2^\dag\tilde B\tilde h_u \\
&\;+\frac{\tilde \gamma_{d}}{\sqrt{2}} H_2\tilde W\tilde h_d
 +\frac{\tilde\gamma_{d}'}{\sqrt{2}} H_2\tilde B\tilde h_d
 +\frac{\tilde \gamma_{u}}{\sqrt{2}}H_1^\dag\tilde W\tilde h_u
 +\frac{\tilde \gamma_{u}'}{\sqrt{2}}H_1^\dag\tilde B\tilde h_u\\
 &\;\hc .
\end{split}
\ee
The gauge symmetries of the general THDM with higgsinos and gauginos further
allow for Yukawa couplings between the higgsinos, right-handed leptons and Higgs
bosons.

If all the couplings allowed by gauge symmetry were actually present (and
sizeable) in the THDM, this would lead to phenomenologically unacceptable rates
of flavour changing neutral currents. However, matching to supersymmetry leads
to strong restrictions on the parameter space as we will now describe in detail.

\subsection{Matching to the MSSM at the scale $M_S$}

We identify $H_1=-i\sigma^2 H_d^*$ and $H_2=H_u$ at the scale $M_S$, where $H_u$
and $H_d$ are the Higgs doublets of the minimal supersymmetric Standard Model.
Tree-level matching at the scale $M_S$ gives
\be\begin{split}\label{quarticmatching}
\lambda_1=&\;\frac{1}{4}\left(g^2+{g'}^2\right),\\
\lambda_2=&\;\frac{1}{4}\left(g^2+{g'}^2\right),\\
\lambda_3=&\;\frac{1}{4}\left(g^2-{g'}^2\right),\\
\lambda_4=&\;-\frac{1}{2}g^2,\\
\lambda_5=&\;\lambda_6=\lambda_7=0\,.
\end{split}
\ee
Here $g\equiv g_2$ and $g'\equiv\sqrt{\frac{3}{5}} g_1$.

The one-loop threshold corrections to these couplings are e.g.~listed in 
Ref.~\cite{Gorbahn:2009pp}. The exact superpartner spectrum at $M_S$ is 
of course unknown, but we use the GUT model of 
Ref.~\cite{Buchmuller:2015jna} as a guidance. It predicts that the 
squark and slepton soft masses are degenerate to leading order at the 
matching scale $M_S$, and that all other soft parameters are generated 
at subleading order. In this case the squark and slepton threshold 
corrections are suppressed not only by a loop factor but also by the 
small ratios $A/M_S$, $\mu/M_S$ and by the near-degeneracy of the 
squarks and sleptons, and their impact on our results is correspondingly 
reduced.

In the following we set these threshold corrections to zero for
definiteness, with the understanding that this is a source of model
dependence. To account for the neglected effects, we will assume a 
conservative 3 GeV uncertainty on $m_h$ in our analysis.


Following the same line of reasoning, we also neglect the higgsino threshold corrections to Eqs.~\eqref{quarticmatching} in the pure THDM case,
and the electroweak gaugino threshold corrections in the case of both the pure 
THDM and the THDM with light higgsinos. 


Note that the tree level matching conditions Eqs.~\eqref{quarticmatching} are
not specific to the UV completion being the MSSM, but apply in any model in
which the quartic scalar potential emerges from the $D$-term potential of an
$N=1$ supersymmetric $\SU{2}\times\U{1}$ theory.

Since we are setting $\lambda_5=\lambda_6=\lambda_7=0$ in our analysis, and
since the Yukawa terms
$H_d^\dag \bar u_R q_L+H_u^\dag \bar d_R q_L+H_u^\dag\bar e_R\ell_L\hc$
are also absent at the matching scale (up to small threshold corrections which
we neglect), our model becomes an effective type-II THDM.

If there are winos or binos in the spectrum, the matching conditions
for their Yukawa couplings at the scale $M_S$ read at the tree-level
\be\begin{split}\label{gauginomatching}
    \tilde g_u=&\;g\,,\\
    \tilde g_d=&\;g\,,\\
    \tilde g_u'=&\;g'\,,\\
    \tilde g_d'=&\;g'\,,\\
    \tilde\gamma_u=\tilde\gamma_d=\tilde\gamma_u'=\tilde\gamma_d'=&\;0\,.
   \end{split}
\ee
 We will again neglect possible effects from small threshold corrections. We
also assume that there is some conserved quantum number (such as $R$-parity or $B-L$)
distinguishing the higgsino from the lepton doublets, such that there are no
Yukawa couplings between the Higgs, the higgsino and the right-handed leptons.

\subsection{Running to the scale $M_t$}

The $\lambda_i$ evolve from $M_S$ down to the electroweak scale according to their
renormalisation group equations. Note that $\lambda_{5,6,7}$,
$\tilde\gamma_{u,d}$, $\tilde\gamma'_{u,d}$, as well as the ``wrong Higgs''
quark and lepton Yukawa couplings, are protected by the symmetries of the effective theory and therefore will
not be generated during the running if they are zero at the matching scale,
which we assume is the case.  We therefore work with all these couplings set to
zero henceforth.

To obtain a scalar potential that is bounded from below, a set of sufficient
conditions on the running scalar couplings is Ref.~\cite{Deshpande:1977rw}
\begin{align}
\lambda_1>&\;0\,,\label{eq:stability1}\\
\lambda_2>&\;0\,,\label{eq:stability2}\\
\lambda_3+\left(\lambda_1\lambda_2\right)^{1/2}>&\;0\,,\label{eq:stability3}\\
\lambda_3+\lambda_4+\left(\lambda_1\lambda_2\right)^{1/2}>&\;0\,.\label{eq:stability4}
\end{align}
Numerically it will turn out that the first three conditions are always satisfied
as a consequence of the supersymmetric matching conditions, while the fourth one
Eq.~\eqref{eq:stability4} may be violated at intermediate scales.

The stability conditions can be relaxed if one allows for additional vacua besides the
electroweak one, and merely imposes that the lifetime of the electroweak vacuum
be $\gtrsim 10^{10}$ years. In that case, assuming that the conditions
(\ref{eq:stability1}--\ref{eq:stability3}) are satisfied, the condition \eqref{eq:stability4}
is replaced by an inequality which should hold at all renormalisation scales $\mu_r$,
\be\label{eq:metastability}
\lambda(\mu_r)\gtrsim -\frac{2.82}{41.1+\log_{10}\frac{\mu_r}{\rm GeV}} \equiv \lambda_{\mathrm{meta}} \,,
\ee
where
\be\label{eq:lambdadef}
\lambda=\frac{4\,(\lambda_1\lambda_2)^{1/2}\,\left(\lambda_3+\lambda_4+(\lambda_1\lambda_2)^{1/2}\right)}{\lambda_1+\lambda_2+2\,(\lambda_1\lambda_2)^{1/2}}\,.
\ee
A derivation of Eq.~\eqref{eq:metastability} is given
in \appendixname~\ref{appendix:metastability}.

In order
to numerically study the running of the parameters in the presence of
the boundary and vacuum stability conditions, we use the spectrum
generator framework FlexibleSUSY 1.2.1 \cite{Athron:2014yba} in
combination with SARAH 4.6.0
\cite{Staub:2008uz,Staub:2010jh,Staub:2013tta}\footnote{The SARAH version
  we use contains an additional bug-fix, which corrects the
  \MSbar--\DRbar~conversion terms in the left- and right-handed
  one-loop fermion self-energies.}. The latter is used to compute the
2-loop renormalisation group equation for the effective field theories.
As a preliminary safety-check, we have compared the expressions obtained from SARAH
with the ones provided by PYR@TE~\cite{Lyonnet:2013dna,Lyonnet:2015jca}, finding
complete agreement.

FlexibleSUSY makes use of 2-loop renormalisation group equations and provides an automatic
matching of the THDM to input parameters at the electroweak scale (we
perform the matching at the scale $M_t$), as described in the following as well as in more detail in
\appendixname~\ref{appendix:matching}. 

\subsection{Matching at the weak scale}

By integrating the 2-loop renormalisation group equations we obtain
the running parameters of the THDM (potentially including higgsinos and
gauginos) at the scale $M_t$, where we match the THDM to
experimentally known input parameters. 
The matching is performed by calculating the \MSbar\ gauge and
Yukawa couplings as well as the VEVs of the THDM from known 
input parameters at the 1- and leading 2-loop level. In particular, at
the tree level, the well-known THDM relations
\be\begin{split}
m_{12}^2=&\;m_A^2\sin\beta\cos\beta\,,\\
m_1^2=&\;m_{12}^2\tan\beta-v^2\left(\lambda_1\cos^2\beta+(\lambda_3+\lambda_4)\sin^2\beta\right)\,,\\
m_2^2=&\;m_{12}^2\cot\beta-v^2\left(\lambda_2\sin^2\beta+(\lambda_3+\lambda_4)\cos^2\beta\right)\,,\\
\end{split}
\ee
allow us to express the entire scalar potential in terms of
$v=\sqrt{v_u^2 + v_d^2}$, the quartic couplings, the pseudoscalar
\MSbar\ Higgs mass $m_A$ and
\be
\tan\beta\equiv\frac{v_2}{v_1}\,.
\ee
More details on the matching procedure at the loop level are given in
\appendixname~\ref{appendix:matching}.

We note that our models have the appealing feature that there are very few parameters
left in the low-energy theory. Since the quartic couplings are essentially determined by the gauge
couplings via the supersymmetric boundary conditions, the only free parameter which directly affects
them is the matching scale $M_S$. Setting $v\approx 174$ GeV implies that, in the pure THDM,
the Higgs mass spectrum is completely determined by the parameters $M_S$, $m_A$ and $\tan\beta$,
one of which can (in principle) be fixed by requiring $M_h=125$ GeV. Moreover,
requiring vacuum stability forces us into the region of rather low $\tan\beta$, and the sensitivity
of the low-energy spectrum to $M_S$ is very mild. This allows us, in principle, to predict a sharp
correlation between $\tan\beta$ and $m_A$. In practice, however, the theory uncertainty
on the calculation of the lightest Higgs mass is still so large that there is still room for
significant variation, as we will detail in the next section.

\subsection{Higgs-mass predictions}

In the THDM with higgsinos, the Higgs masses receive loop corrections
from char\-gi\-nos and neutralinos and hence
depend on the higgsino mass parameter $\mu$. This leads to correlations between the Higgs and neutralino and
chargino masses which are in principle testable at colliders. In the THDM with higgsinos and gauginos, the
Higgs masses depend on all the chargino and neutralino masses, and may in addition be affected by two-loop
corrections from the gluino. This will also become evident in the next section.

  We calculate the CP-even Higgs pole masses by numerically finding
  the two eigenvalues $M_{h,H}^2$ of the one-loop-corrected mass
  matrix
  \begin{align}
    \underline{M}_{h,\text{1L}}^2 = \underline{M}_{h}^2
       - \re\Sigma_h(p^2=M_{h,H}^2,\mu_r=M_t).
    \label{eq:loop-corrected_Higgs_mass_matrix}
  \end{align}
  Here, $\underline{M}_{h}^2$ denotes the CP-even Higgs mass matrix
  expressed in terms of the \MSbar\ parameters at the scale $\mu_r =
  M_t$ and $\Sigma_h(p^2,\mu_r)$ is the \MSbar\ renormalised CP-even
  Higgs one-loop self-energy matrix, where the Higgs fields at the
  external legs are taken to be the Higgs gauge eigenstates.  Since
  the Higgs self-energy has to be evaluated at the momenta $p^2 =
  M_{h,H}^2$, where $M_{h,H}^2$ are the eigenvalues of
  $\underline{M}_{h,\text{1L}}^2$,
  Eq.~\eqref{eq:loop-corrected_Higgs_mass_matrix} is solved
  iteratively.

\section{Results}\label{sec:Results}

\begin{figure}
 \begin{center}
 \begin{tabular}{ll}
  \includegraphics[width=0.5\textwidth]{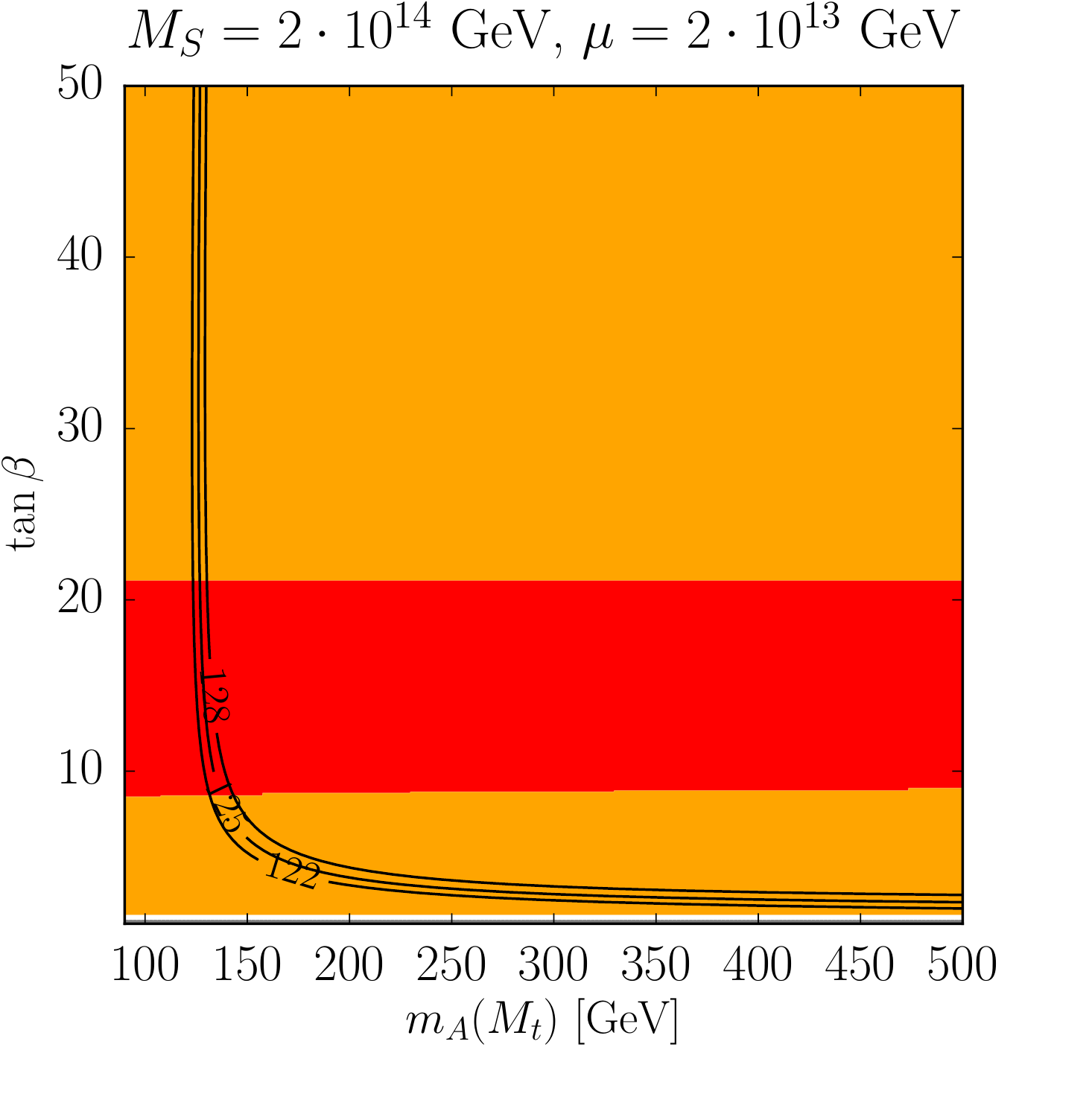}
&
  \includegraphics[width=0.5\textwidth]{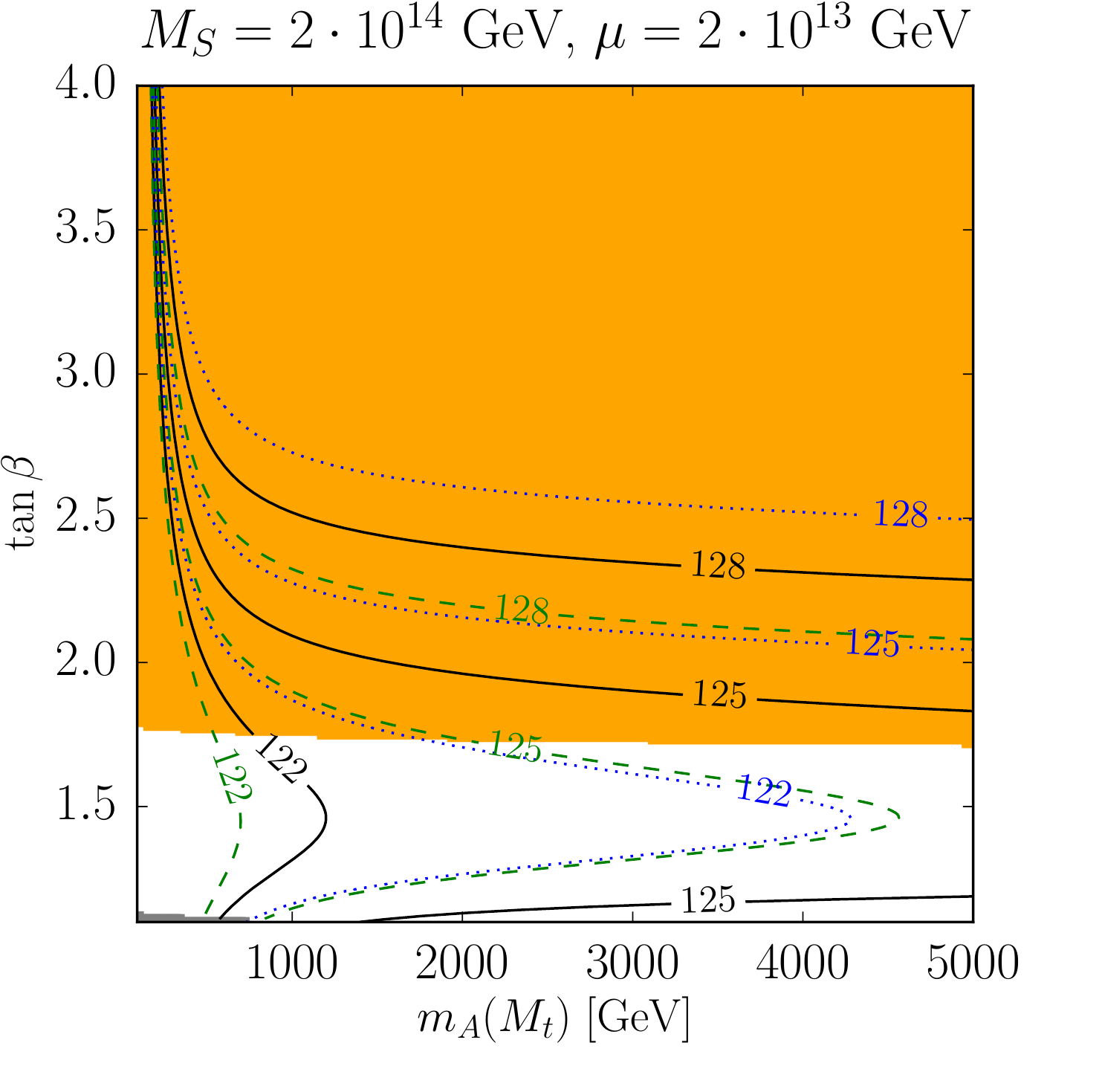}\\
  \includegraphics[width=.5\textwidth]{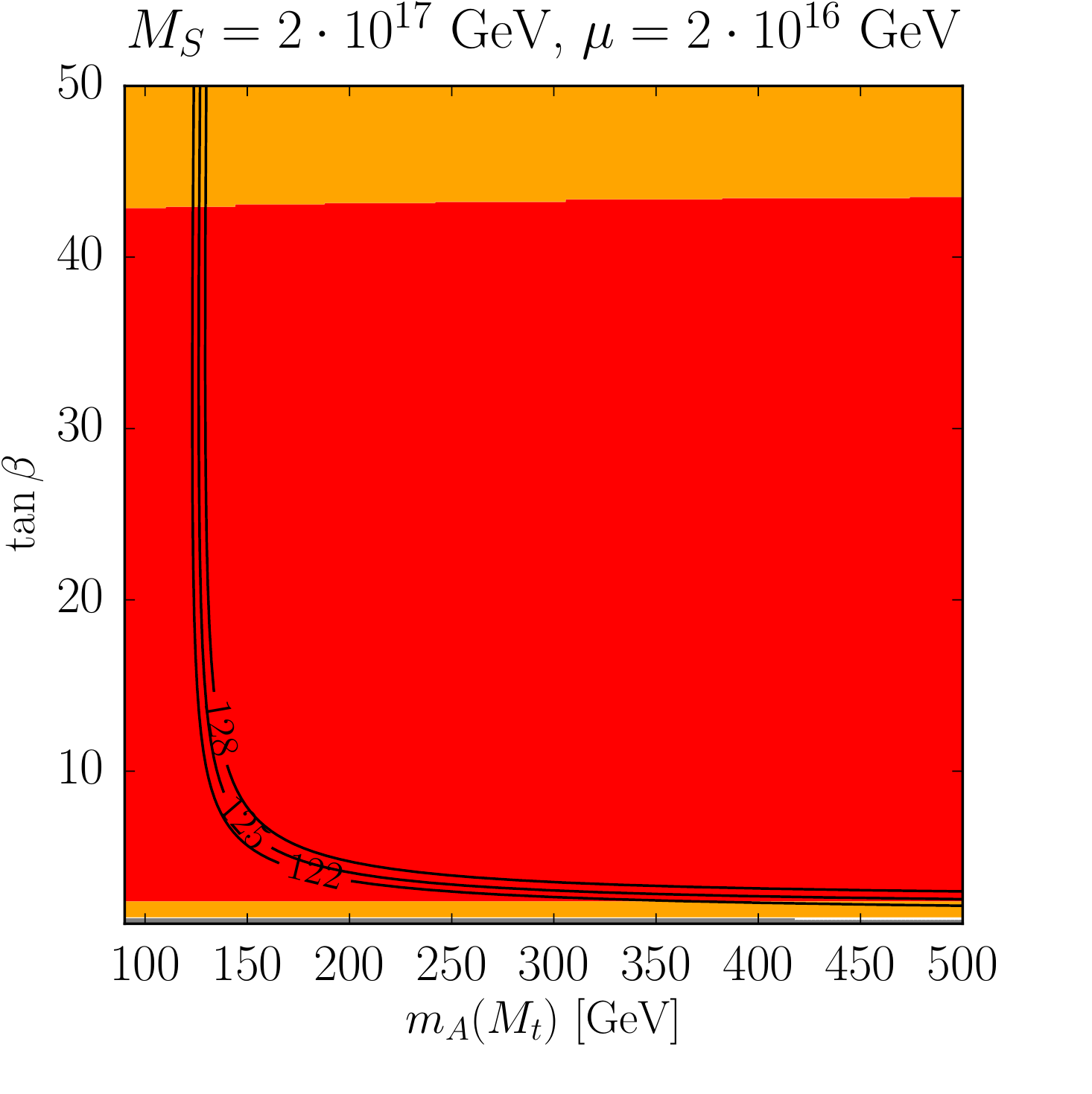}
&
  \includegraphics[width=.5\textwidth]{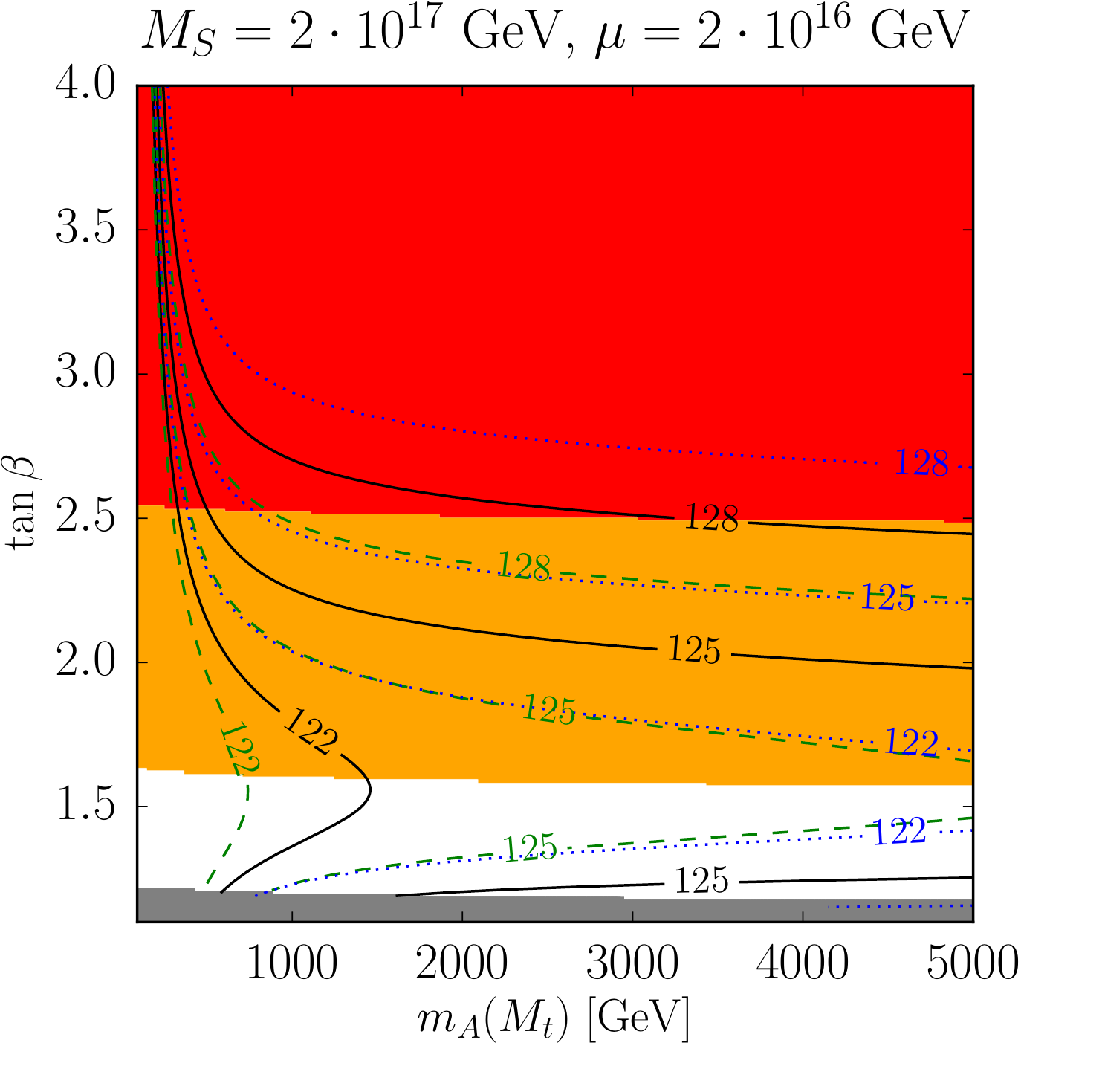}
 \end{tabular}
 \includegraphics[width=0.9\textwidth]{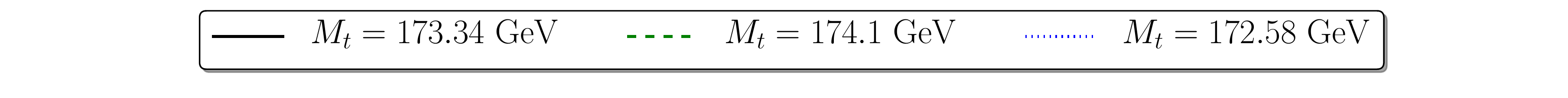}
  \end{center}
\caption{Contours of the lightest Higgs mass $M_h$ in the $m_A(M_t)$ --
$\tan\beta$ plane in the pure THDM for $M_S=2\cdot 10^{14}$ GeV (top row)
and $M_S=2\cdot 10^{17}$ GeV (bottom row). The Higgs mass prediction is computed for $M_t= 173.34 \pm 0.76$ GeV (solid black,
dashed green and dotted blue).
Left: full range of $\tan\beta$, low $m_A(M_t)$; right: region of low $\tan\beta$, large $m_A(M_t)$. Unshaded regions are allowed by vacuum stability. In the orange region, the electroweak
vacuum is unstable but its lifetime is larger than the age of the universe. Red regions are excluded by vacuum stability. Grey
regions are uncalculable because perturbative control is lost.
}
\label{fig:mh-mA-cs-tanb-MS-THDM}
\end{figure}

\begin{figure}
 \begin{center}
 \begin{tabular}{ccc}
  \includegraphics[width=.5\textwidth]{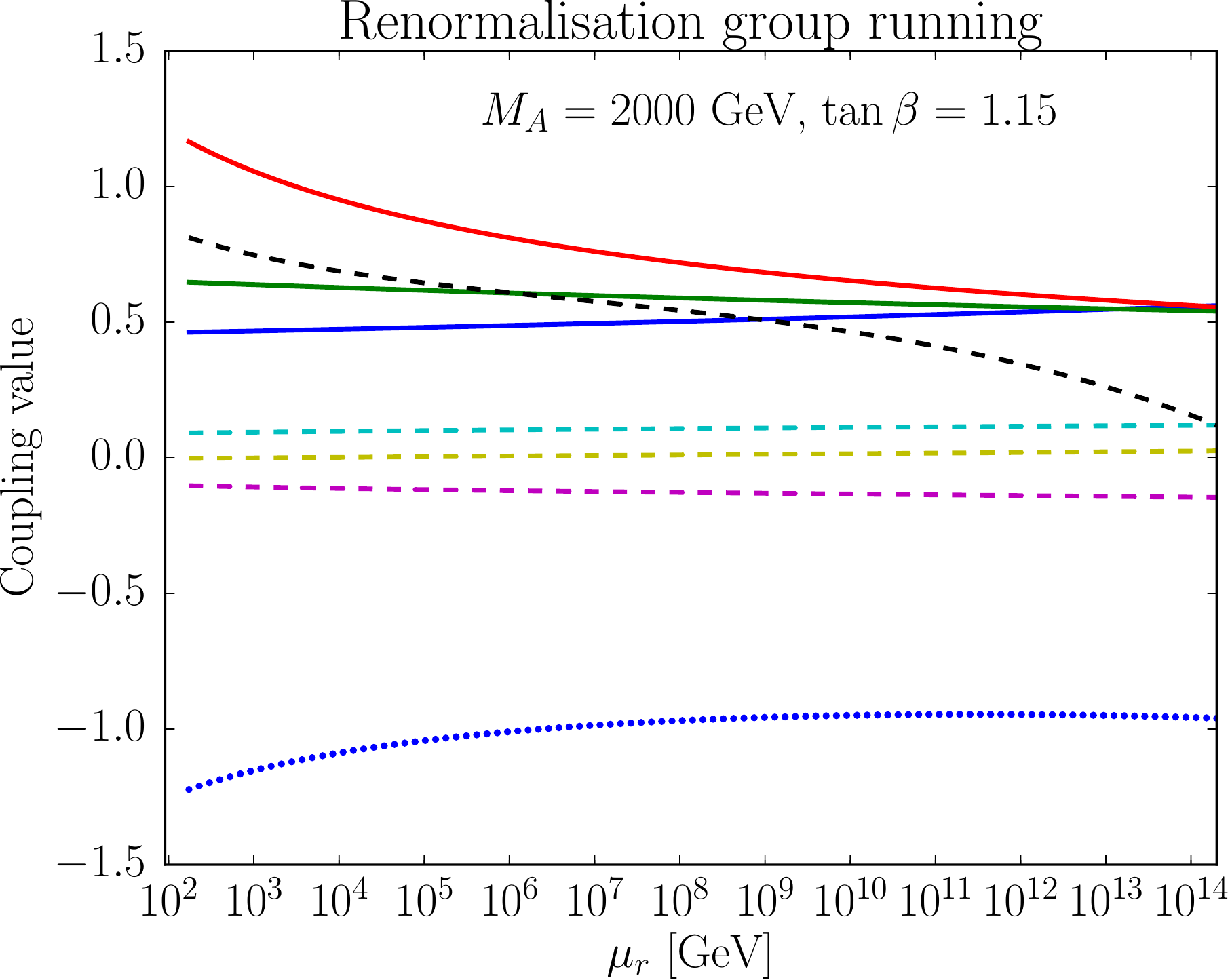}
  &&
  \includegraphics[width=.5\textwidth]{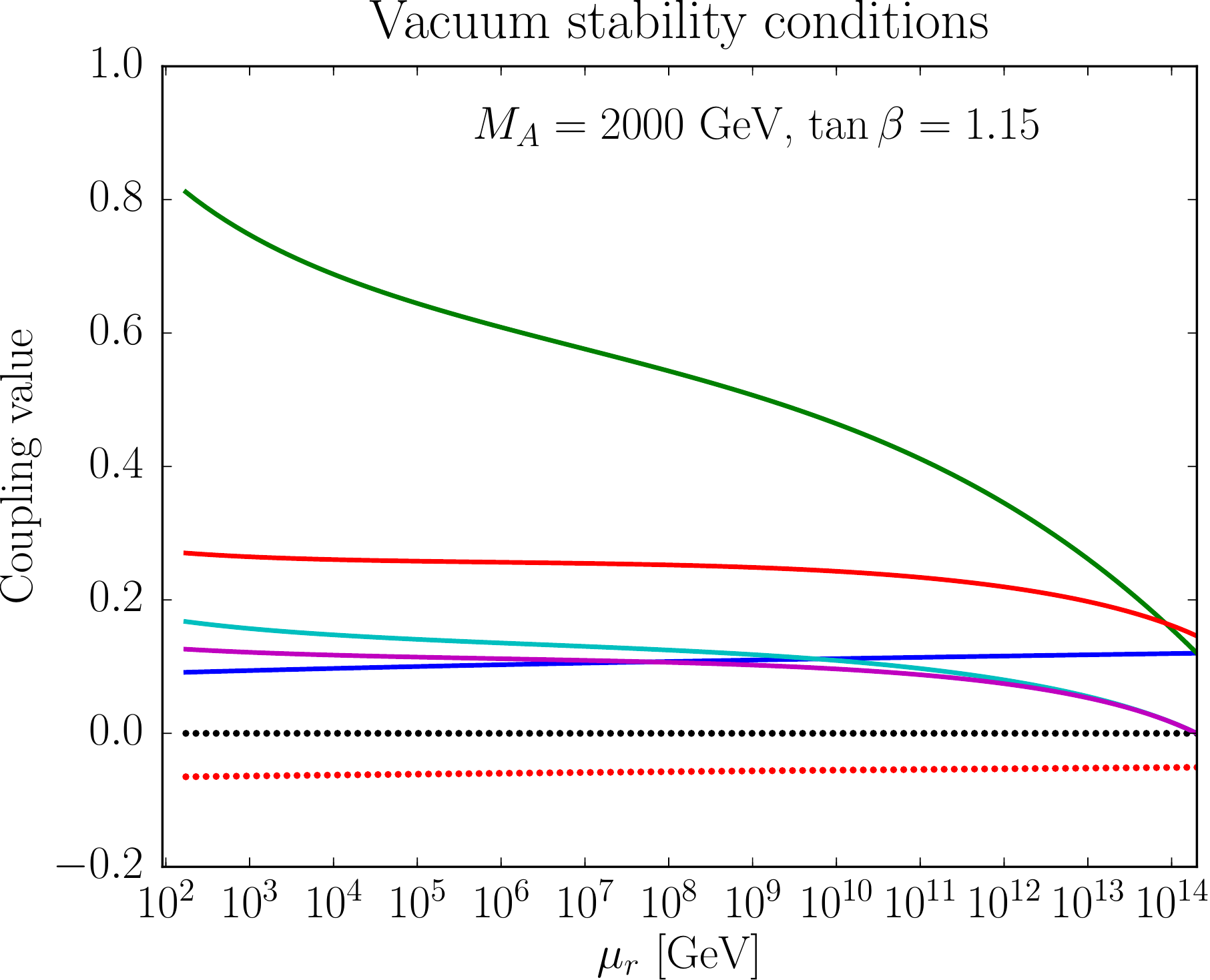}\\[1cm]
  \includegraphics[width=.5\textwidth]{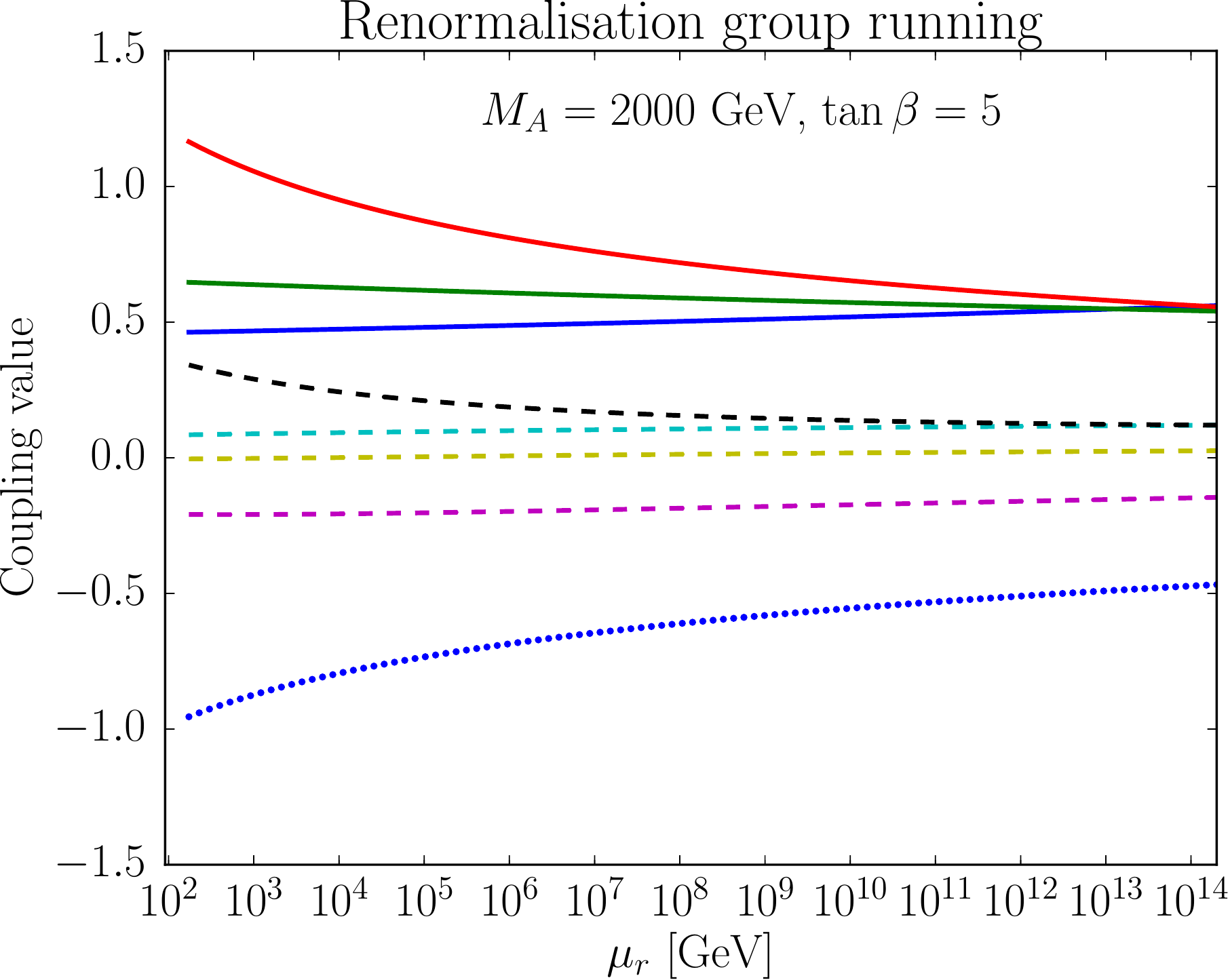}
  &&
  \includegraphics[width=.5\textwidth]{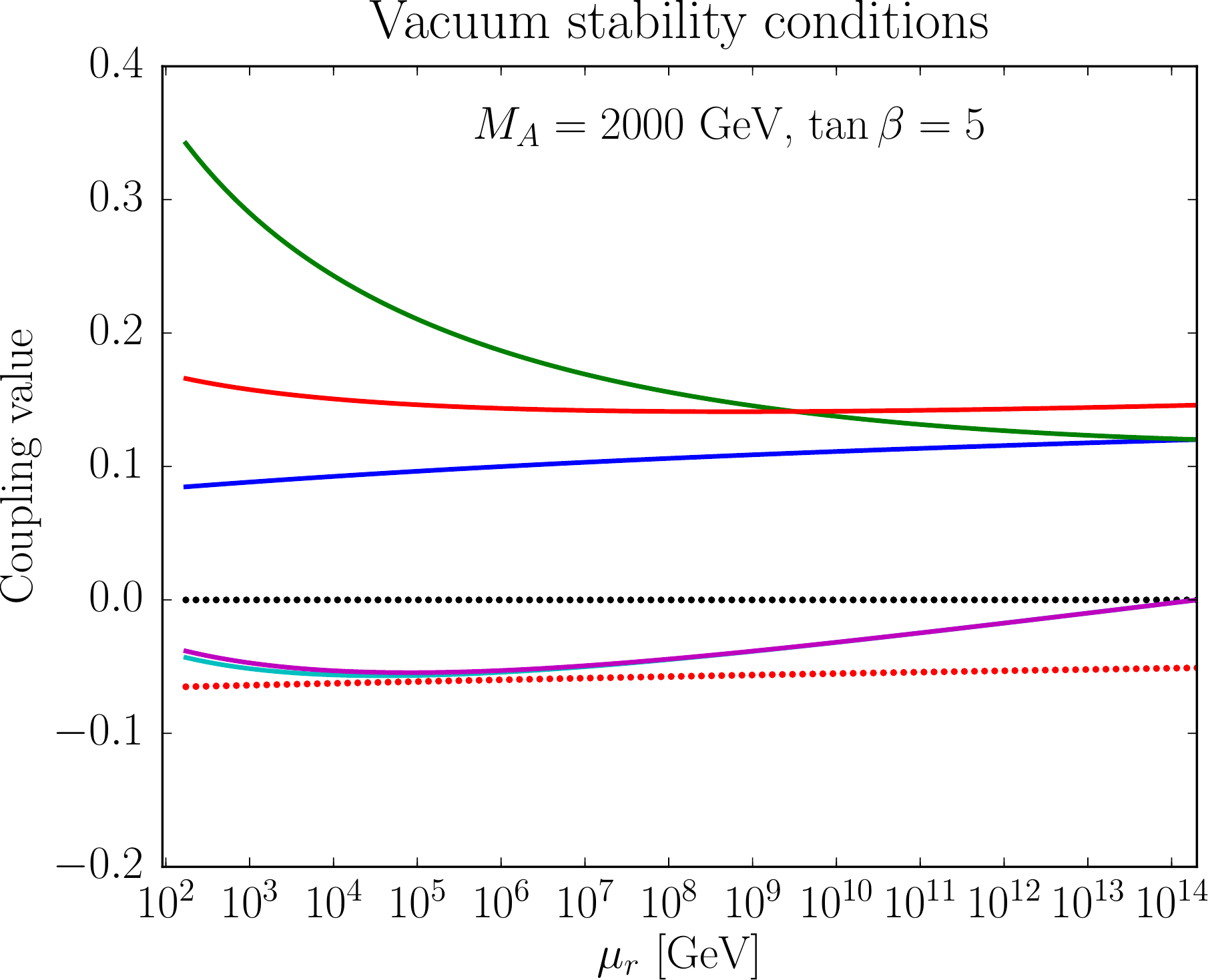}\\[1cm]
  \includegraphics[width=.5\textwidth]{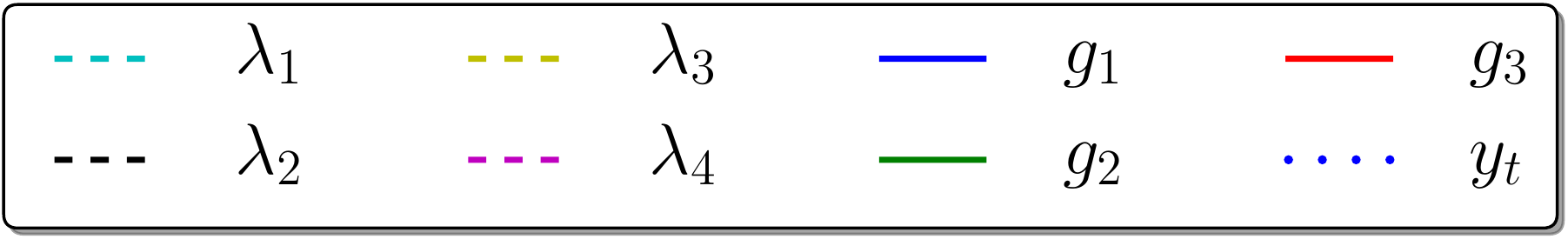}
  &&
  \includegraphics[width=.5\textwidth]{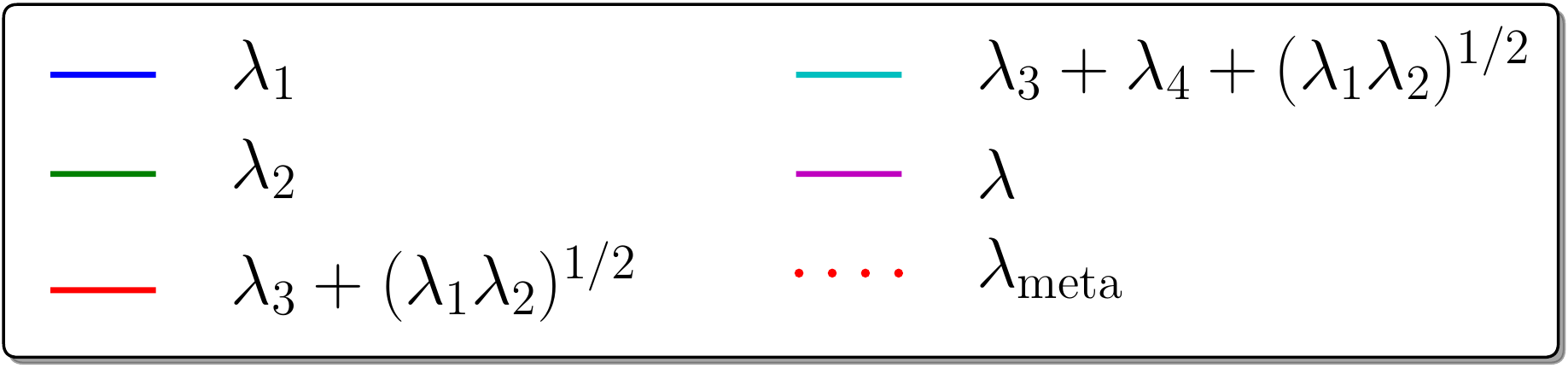}
 \end{tabular}
  \end{center}
  \caption{Renormalisation group running of dimensionless parameters
    (left column) and the vacuum stability conditions (right column),
    in the THDM for $M_S=2\cdot 10^{14}$~GeV, 
    for two different points characterised by a stable
    electroweak vacuum (top row) and metastable behaviour (bottom row).
    $\mu_r$ denotes the renormalisation scale. $\lambda$ and $\lambda_{\rm meta}$ 
    are defined in Eqs.~\eqref{eq:metastability} and \eqref{eq:lambdadef}.}
\label{fig:RGE-flow-mA-1000}
\end{figure}

\subsection{The pure THDM}

The low-energy parameter space is strongly constrained by vacuum (meta)stability, by requiring the lightest
Higgs boson mass to be $125$ GeV, 
and by the experimental bounds 
from the measurement of ${\rm BR}(b\into s\gamma)$~\cite{Amhis:2014hma}
and the limits from
the LHC searches for 
$H,A\into\tau\tau$~\cite{Aad:2014vgg,CMS:2015mca,ATLASHtautau13TeV}.
In the top row of Fig.~\ref{fig:mh-mA-cs-tanb-MS-THDM} we show contours of the lightest Higgs mass as a function of
$M_A$ and $\tan\beta$ for a SUSY breaking scale $M_S=2\cdot 10^{14}$ GeV. The vacuum is absolutely stable
only in the white unshaded region at low $\tan\beta$. It is metastable in the bulk of the parameter space
(orange regions), and unstable in the red region of intermediate $\tan\beta$.

We remark that including high-scale one-loop threshold corrections from
heavy higgsinos, which we have neglected
 in generating these plots, can have a significant impact on the large $\tan\beta$ region. For example,
choosing $\mu=0.1~M_S$ somewhat lowers the upper boundary of the unstable region and opens up a new stable region
around $\tan\beta = 30$. However, the constraint $M_h=125$ GeV enforces $M_A\lesssim 200$ GeV at large
$\tan\beta$, and this parameter region is excluded by 
the constraint on the charged Higgs boson mass in a THDM 
from the measurement of ${\rm BR}(b\into s\gamma)$
(since the charged Higgs is similarly light as the pseudoscalar) and by 
the limits from the LHC searches for 
$H,A\into\tau\tau$.
Thus, including
or neglecting these threshold corrections only affects a parameter
region which is phenomenologically disfavoured
anyway.

Note that absolute vacuum stability forces one into the low 
$\tan\beta$ region, $\tan\beta\lesssim 1.8$, with pseudoscalar
Higgs masses exceeding a TeV for $M_h = 125$~GeV and the central value
of $M_t$. 
By contrast, when allowing for the vacuum to be metastable, 
the most severe constraint
on $M_A$ comes from the measurement of ${\rm BR}(b\into s\gamma)$,
which together with the requirement that $M_h$ should be close to 125~GeV
still points to somewhat small $\tan\beta$ values, $\tan\beta \lsim 5$.

For comparison, we also show the case of a higher SUSY breaking scale $M_S=2\cdot 10^{17}$ GeV in the bottom row of
Fig.~\ref{fig:mh-mA-cs-tanb-MS-THDM}. This scale, an order of magnitude below $M_{\rm Planck}$, is about the
highest for which the matching to a weakly coupled four-dimensional supersymmetric field theory can be justified.
While the qualitative behaviour in the plane is the same as for the lower SUSY breaking scale case, we observe that
a large part of the formerly
metastable region is now unstable. Concerning the higgsino one-loop threshold corrections, similar remarks
as above apply\footnote{Note that for part of the parameter space
  considered in Ref.~\cite{Lee:2015uza}, $m_A = 200$~GeV and $M_S$ of the
  order of the grand unification scale, the electroweak vacuum is
  either metastable or unstable.}.

 In order to understand why the THDM allows a matching to the
supersymmetric standard model at very high scales one has to study the
renormalisation group flow of the quartic couplings. This is shown in
Fig.~\ref{fig:RGE-flow-mA-1000}  for $M_S=2\cdot 10^{14}$ GeV for two values
of $\tan\beta$. For small values of $\tan\beta$ the absolute value of
the top-quark Yukawa
coupling is large in the IR. This drives the coupling $\lambda_2$ also
to large values in the IR. In the UV, at $M_S$, all quartic couplings
are determined by the gauge couplings, which approximately unify in
the THDM. Due to the boundary conditions the coupling $\lambda_4$ is
negative at $M_S$. Hence the condition
$\lambda_3+\lambda_4+\sqrt{\lambda_1\lambda_2} > 0$
is the most stringent stability constraint. As
Fig.~\ref{fig:RGE-flow-mA-1000} shows, for $\tan\beta=1.15$ the
coupling $\lambda_2$ is sufficiently large such that
$\sqrt{\lambda_1\lambda_2}$ can compensate the negative
$\lambda_4$. For $\tan\beta=5$ this is no longer the case, and only
the weaker metastability condition is satisfied.

\begin{figure}
  \begin{center}
    \begin{tabular}{ccc}
  \includegraphics[width=.5\textwidth]{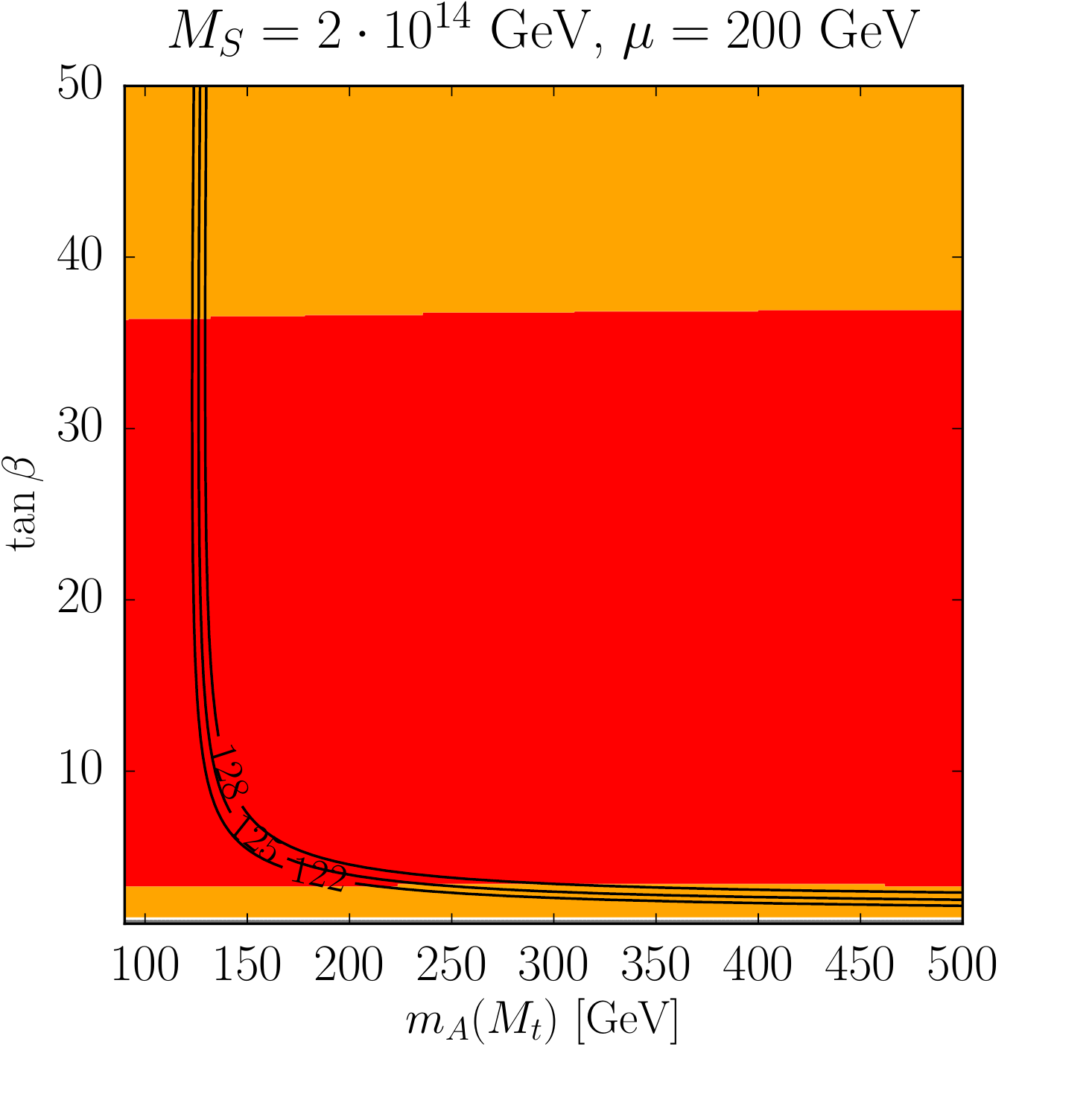}
  &&
  \includegraphics[width=.5\textwidth]{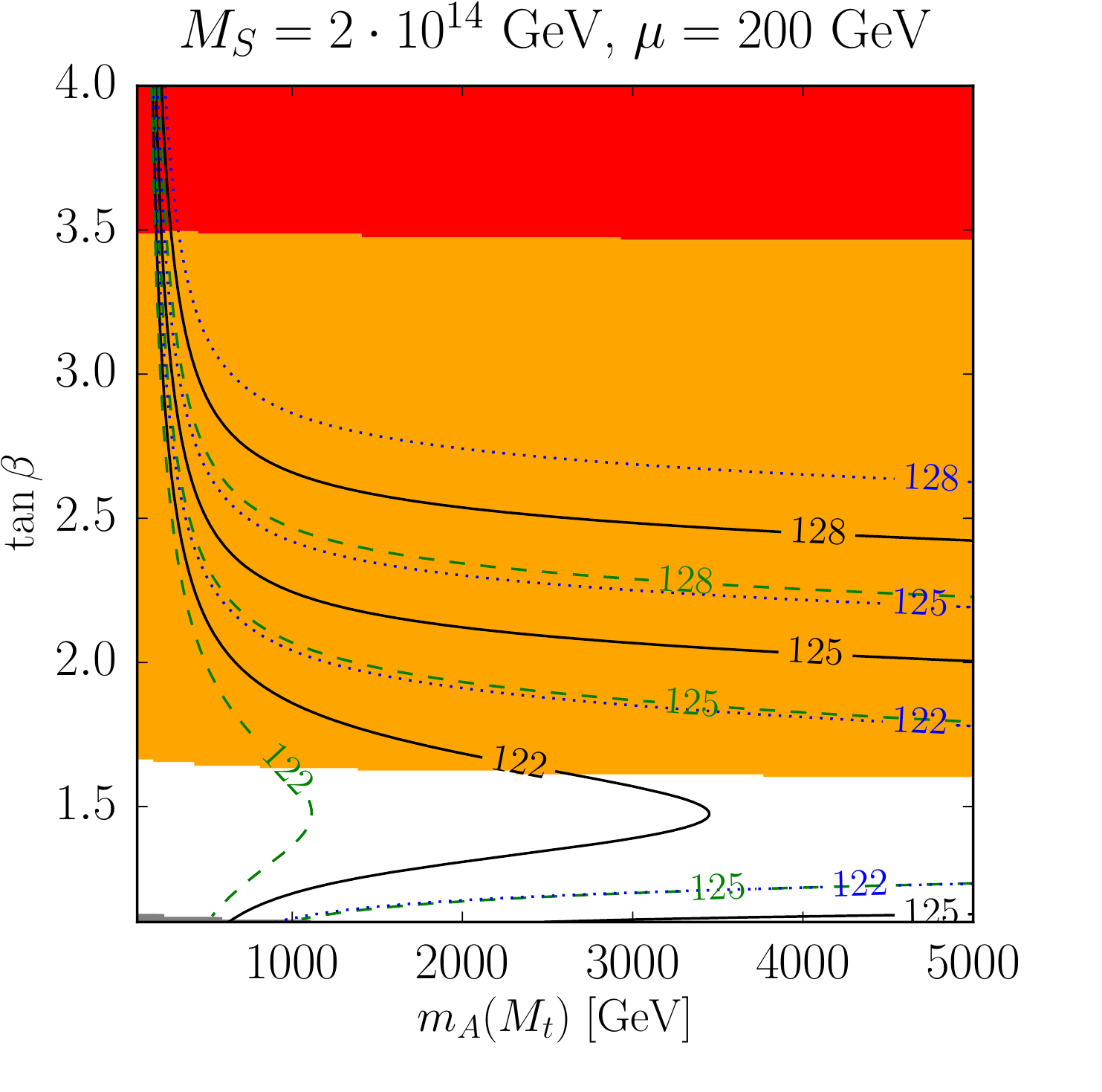}
  \\
  \includegraphics[width=.5\textwidth]{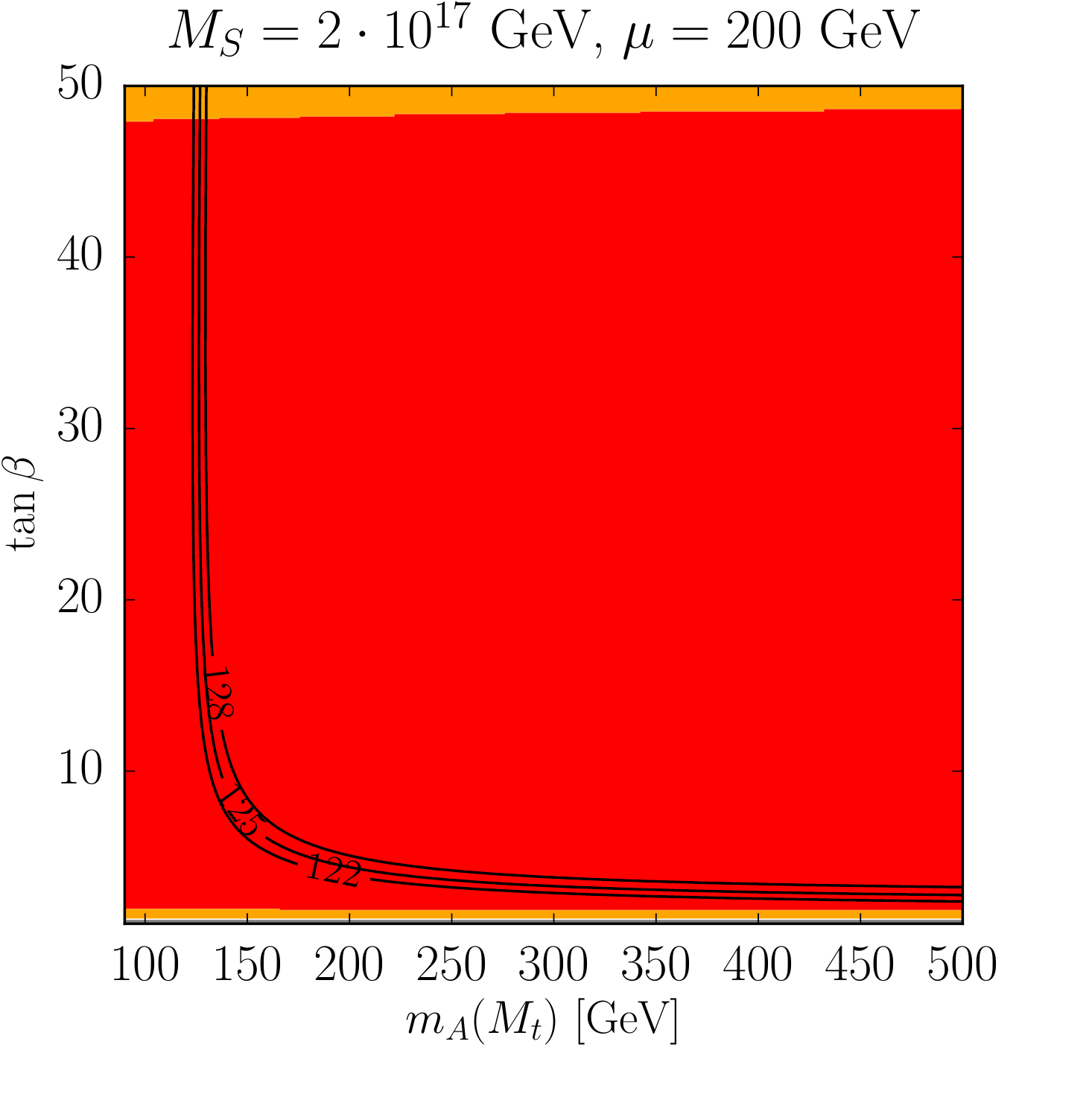}
  &&
  \includegraphics[width=.5\textwidth]{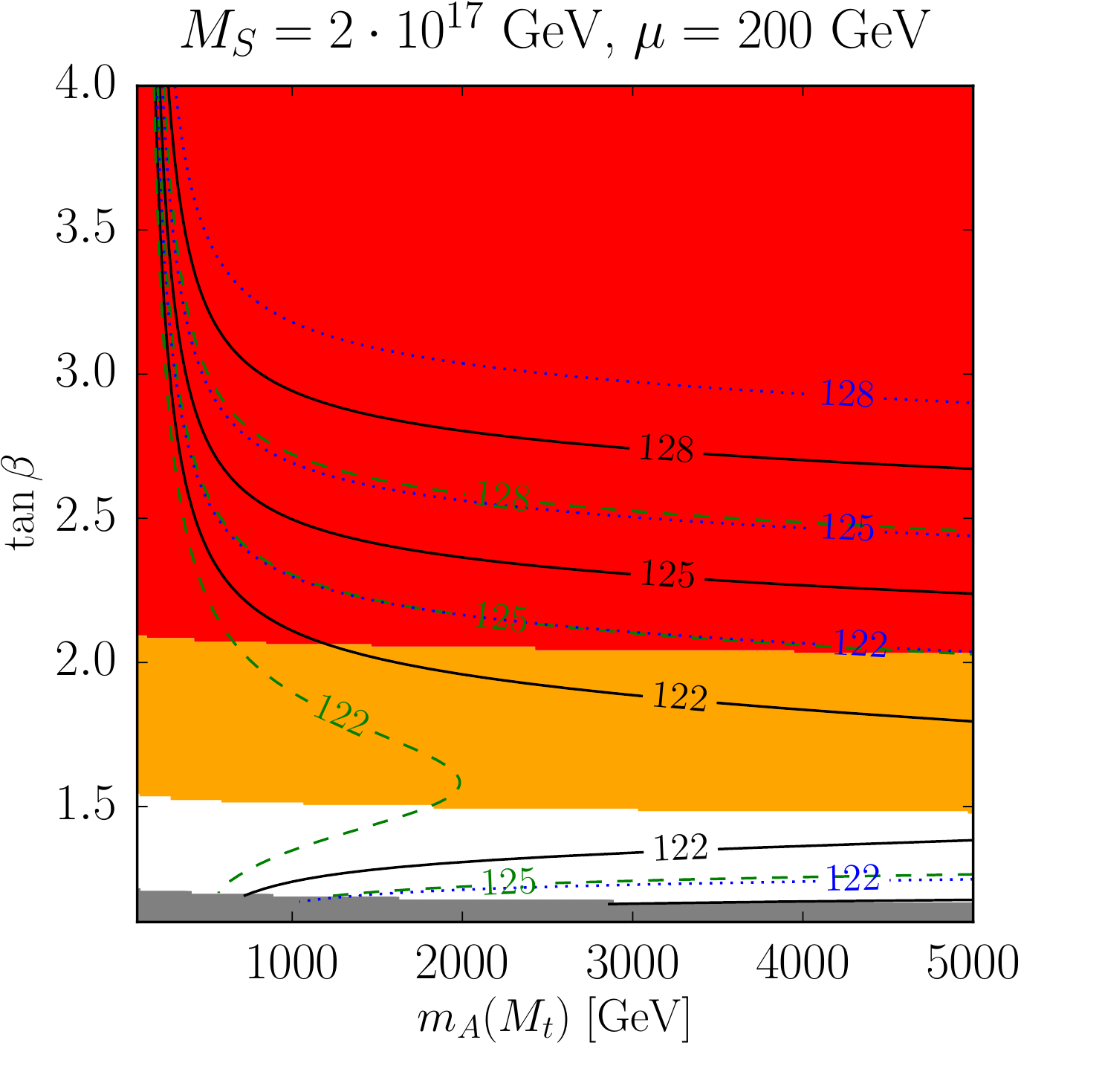}
    \end{tabular}
 \includegraphics[width=0.9\textwidth]{mtop_mhiggs_legend.pdf}
  \end{center}
\caption{Contours of the lightest Higgs mass $M_h$ in the $m_A(M_t)$ --
$\tan\beta$ plane for the case where the spectrum at the electroweak
scale consists of the THDM with higgsinos, with $\mu=200$ GeV,
for $M_S=2\cdot 10^{14}$ GeV (top row) and $M_S=2\cdot 10^{17}$ GeV (bottom row).
The Higgs mass prediction is computed for $M_t= 173.34 \pm 0.76$ GeV (solid black, dashed green and dotted blue).
Left: full range of $\tan\beta$, low $m_A(M_t)$; right: region of low $\tan\beta$, large $m_A(M_t)$. Unshaded regions are allowed by vacuum stability. In the orange region, the electroweak
vacuum is unstable but its lifetime is larger than the age of the universe. Red regions are excluded by vacuum stability. Grey
regions are uncalculable because perturbative control is lost.
}
\label{fig:mh-mA-cs-tanb-MS-HTHDM}
\end{figure}

\subsection{The THDM with higgsinos}

In the case that the gauginos, squarks and sleptons are decoupled at the scale 
$M_S$, while the Higgs bosons of the THDM and their superpartners 
have masses at the electroweak scale, the low-energy mass spectrum depends
on the additional parameter $\mu$. Fig.~\ref{fig:mh-mA-cs-tanb-MS-HTHDM}
shows the results for $\mu=200$ GeV; the picture is qualitatively very
similar for $\mu=2000$ GeV. Already at $M_S=2\cdot 10^{14}$ GeV
a wide range of $\tan\beta$ values is now excluded because the vacuum 
is unstable. For a metastable vacuum the 
requirement that $M_h$ should be close to 125~GeV favours somewhat higher
$M_A$ values than for the pure THDM, in accordance with the 
constraint from the measurement of ${\rm BR}(b\into s\gamma)$.
An absolutely stable region remains at small values of $\tan\beta$,
favouring somewhat higher $M_A$ values than in the pure THDM case. For a
higher SUSY breaking scale $M_S=2\cdot 10^{17}$ GeV the parameter
space is even more constrained.

It is important to notice that the existence of a stable region at
small $\tan\beta$ imposes no constraints on the parameter
$\mu$. Hence, a scenario where at the weak scale the particle content
of the Standard Model is supplemented by the Higgs bosons of a second
doublet at about a TeV
and light neutral and charged higgsinos is fully compatible
with the matching to a supersymmetric UV completion at the grand
unification scale. A discovery of light higgsinos at the LHC could therefore
be interpreted as a possible hint for a supersymmetric UV completion 
at the grand unification scale.

\subsection{The THDM with split supersymmetry}

When retaining the full gaugino spectrum of the MSSM as well as its complete Higgs sector as the light degrees of freedom, this
particle content has the appealing feature that the gauge couplings approximately unify at the scale $M_{\rm GUT}=2\cdot 10^{16}$ GeV.
The best-motivated choice for the matching scale in this case
is therefore $M_S=M_{\rm GUT}$.

The low-energy spectrum now depends 
on the gaugino masses $M_{1,2,3}$ as
well as on $\mu$. For simplicity we choose a common low-scale value $M_1=M_2=\mu$ for the electroweak superpartner masses,
while keeping $M_3 = 2000$ GeV to avoid experimental limits from LHC Run 1.
(Alternatively we could have imposed gaugino mass unification at $M_{\rm GUT}$, which leads to very similar results for
a low-scale value of $M_2$ equal to $\mu$.)
The Higgs sector is affected by the gluino only through two-loop effects,
and therefore is not very sensitive
to the precise value of $M_3$, given that the squarks are
decoupled.  We can therefore assume that the gluino is sufficiently heavy to
have escaped detection at the LHC so far.

\begin{figure}
  \begin{center}
    \begin{tabular}{lll}
  $M_S = 2 \cdot 10^{16}$ GeV, $\mu = M_{1,2,3} = 2000$ GeV && $M_S = 2 \cdot 10^{16}$ GeV, $\mu = M_{1,2,3} = 2000$ GeV\\
  \includegraphics[height=.47\textwidth]{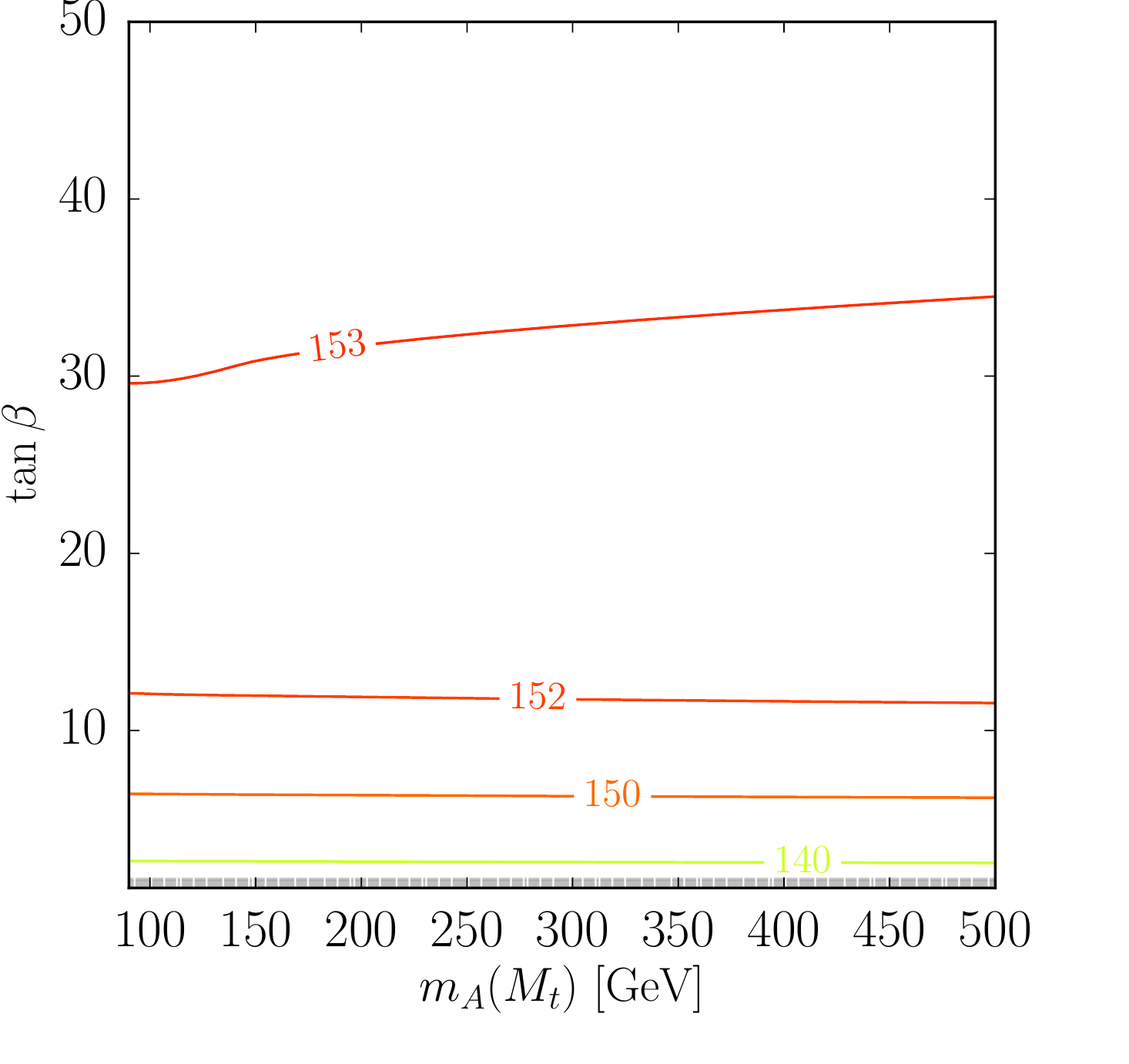}
  &&
  \includegraphics[height=.47\textwidth]{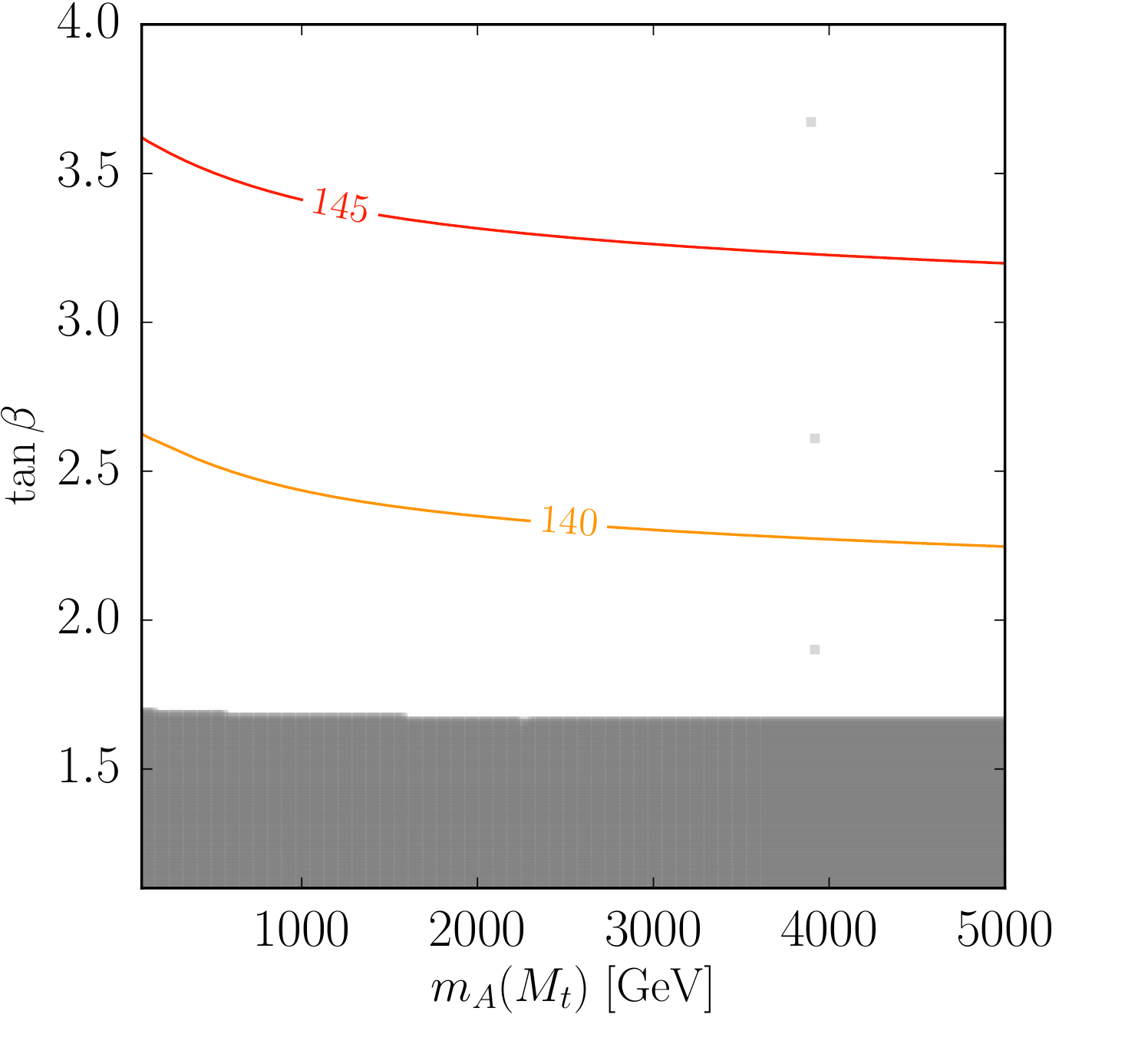}\\
  $\qquad\qquad\quad M_S = 2 \cdot 10^{16}$ GeV && $\qquad\qquad\quad  M_S = 2 \cdot 10^{16}$ GeV \\
  $\quad\mu = M_{1,2} = 200$ GeV, $M_3 = 2000$ GeV && $\quad\mu = M_{1,2} = 200$ GeV, $M_3 = 2000$ GeV \\
  \includegraphics[height=.47\textwidth]{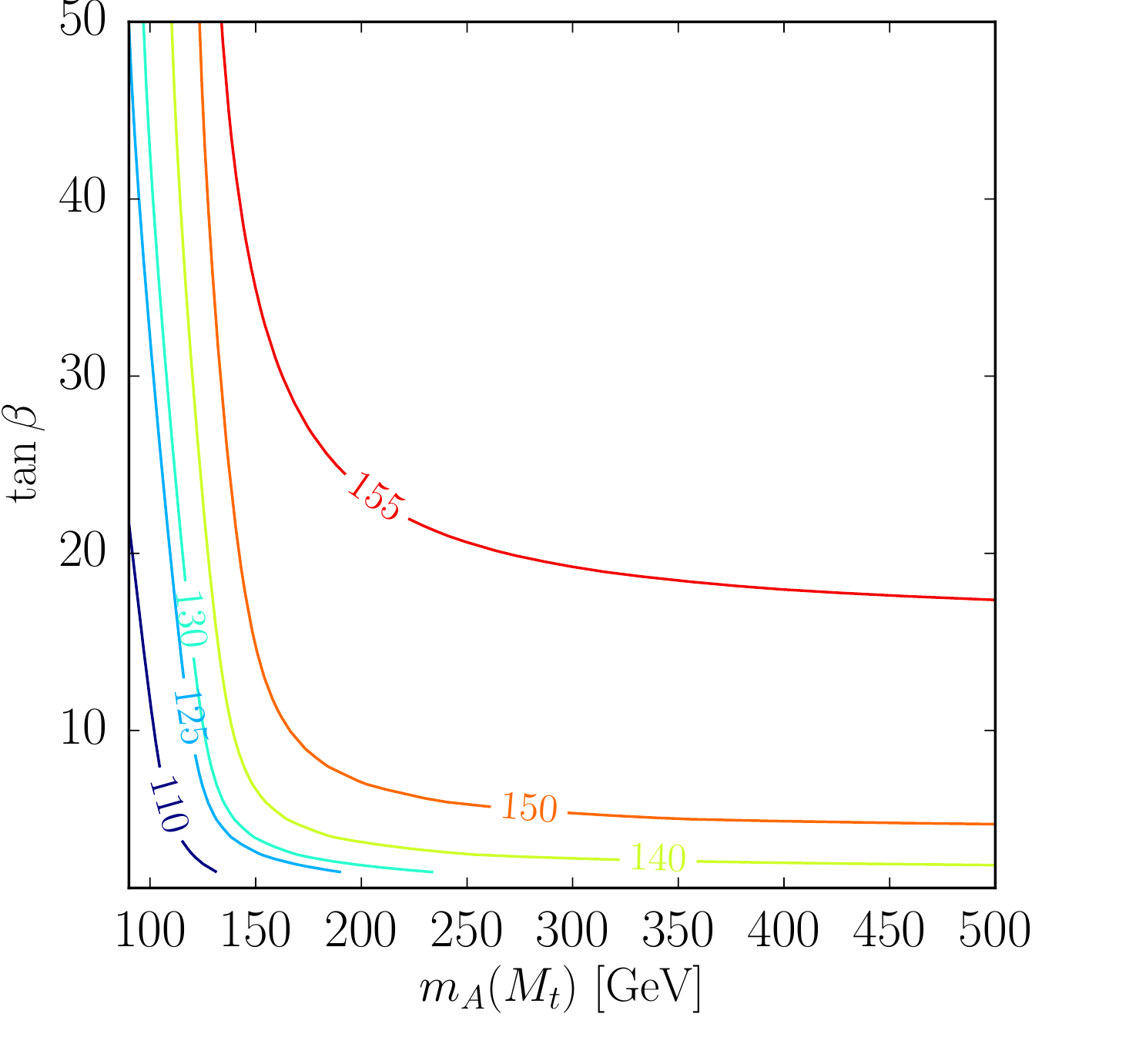}
  &&
  \includegraphics[height=.47\textwidth]{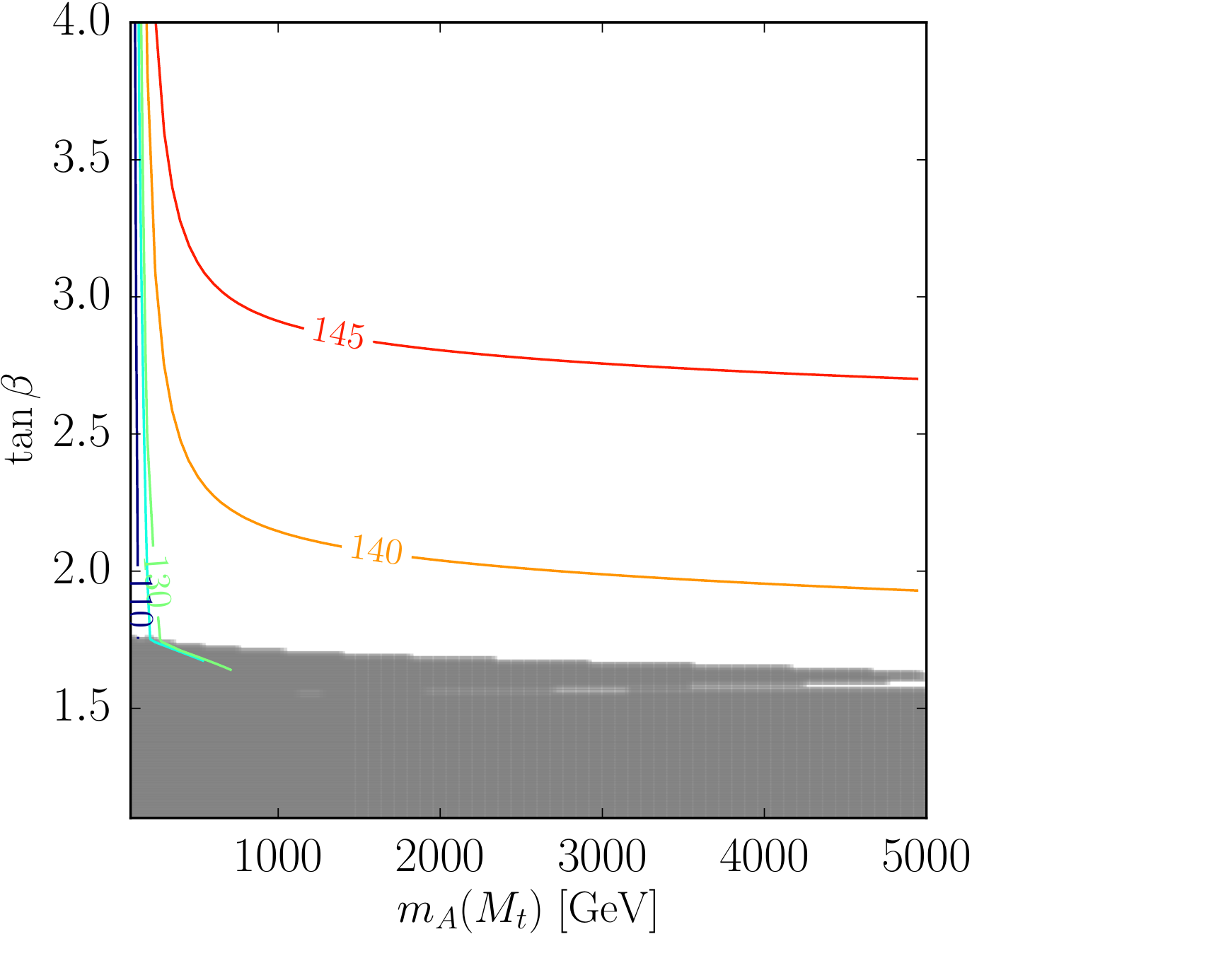}
    \end{tabular}
  \end{center}
\caption{Contours of the lightest Higgs mass $M_h$ in the $m_A(M_t)$ --
$\tan\beta$ plane for the case where the spectrum at the electroweak
scale consists of the THDM with gauginos and higgsinos
(split-supersymmetry) for $M_S=2\cdot 10^{16}$ GeV,
with $\mu=2000$ GeV (top row)
         and $\mu=200$ GeV (bottom row).
Left: full range of $\tan\beta$, low $M_A$; right: region of low $\tan\beta$, large $M_A$. Unshaded white regions are allowed by vacuum stability.
Grey regions are uncalculable because perturbative control is lost.
}
\label{fig:mh-mA-cs-tanb-MS-splitTHDM}
\end{figure}

We find that in the case of light gauginos the vacuum stability
conditions are always satisfied and therefore imply no constraint on
$\tan\beta$. 
As shown in Fig.~\ref{fig:mh-mA-cs-tanb-MS-splitTHDM},
however, a Higgs mass consistent with observation can
only be obtained for small values of $M_A$ which are essentially 
excluded by the
constraint from the measurement of ${\rm BR}(b\into s\gamma)$ in this
scenario.
Hence, the extrapolation of the THDM with
light higgsinos and gauginos up to the grand unification scale is
not compatible with the measured value of the Higgs boson mass.

\section{Summary and outlook}

We have studied the matching of the Standard Model, supplemented by a
second Higgs doublet, with or without additional higgsinos and
gauginos, to the supersymmetric standard model at high scales close to
the GUT scale. A supersymmetric ultraviolet completion of the Standard
Model is strongly motivated by unified theories, in particular string
theory.

The extrapolation of the Standard Model to high scales is severely
constrained by the necessary requirement of stability or metastability
of the electroweak vacuum. In the Standard Model a matching to its
supersymmetric extension at the GUT scale is not possible for the
measured mass of the Higgs boson. On the contrary, as we have shown,
a matching consistent with vacuum stability is possible for two-Higgs-doublet
models. For small values of $\tan\beta$ the large top-quark
Yukawa coupling drives one of the quartic Higgs couplings to large
values in the IR. As a consequence, all vacuum stability conditions can be
satisfied.

The matching of the pure THDM to its supersymmetric extension at high
scales implies a lower bound on the additional Higgs boson masses of
about a TeV. This bound shows a significant sensitivity on the
remaining theoretical uncertainties induced by the experimental error of the
mass of the top quark and from unknown higher-order corrections.
In case of light higgsinos the lower bound is slightly more
stringent than for the case of the pure THDM. 
Because of this preference for low values of $\tan\beta$ and relatively high
values of $M_A$,
the discovery of additional Higgs bosons at the LHC
appears challenging in this scenario. 
In principle, smaller pseudoscalar masses can be
consistent with a metastable electroweak vacuum. But these values of
$M_A$ are already essentially excluded by the constraints from
rare processes. Finally, in the case of both higgsinos
and gauginos at the TeV scale the vacuum stability conditions are
always fulfilled, but a Higgs mass of $125$~GeV implies values of
$M_A$ that are incompatible with low energy measurements.

It is remarkable that the extrapolation of two-Higgs-doublet models
to the GUT scale implies essentially
no constraints on the masses of light neutral
and charged higgsinos, the superpartners of the two Higgs
doublets. Hence, a discovery of just light higgsinos at
the LHC could be interpreted as a possible hint for a 
supersymmetric UV completion 
at the grand unification scale.

The Standard Model requires fine-tuning of the cosmological constant
and the Higgs mass. In two-Higgs-doublet models also the mass term of
the second Higgs doublet has to be fine-tuned. This situation is
unsatisfactory. It is conceivable that an explanation of this puzzle
will eventually be provided by the UV completion.

\subsection*{Acknowledgements}

This work has been supported by the German Science Foundation (DFG) within
the Collaborative Research Center 676 ``Particles, Strings and the Early
Universe''. This research was also supported in part by the European 
Commission through
the ``HiggsTools'' Initial Training Network PITN-GA-2012-316704.
We thank J.~Bernon, S.~Kraml and S.~Shirai for useful discussions.
FB thanks the DESY theory group for hospitality.


\appendix
\section*{Appendices}
\addcontentsline{toc}{section}{Appendices}       
\renewcommand{\thesubsection}{\Alph{subsection}} 

\subsection{Details on the matching at the weak scale}
\label{appendix:matching}

In the following the applied procedure for the matching at the weak
scale is described. The matching is performed at the scale $M_t$.

The \MSbar\ gauge couplings $g_i(M_t)$ of the THDM are calculated as
\begin{align}
  g_1(M_t) &= \sqrt{\frac{5}{3}} \frac{\sqrt{4\pi\alpha_{\text{em}}^{\THDM}(M_t)}}{\cos\theta_W} \,, \\
  g_2(M_t) &= \frac{\sqrt{4\pi\alpha_{\text{em}}^{\THDM}(M_t)}}{\sin\theta_W} \,, \\
  g_3(M_t) &= \sqrt{4\pi\alpha_{\text{s}}^{\THDM}(M_t)} \,,
\end{align}
where $\alpha_{\text{em}}^{\THDM}$ and $\alpha_{\text{s}}^{\THDM}$
denote the electromagnetic and strong coupling constants of the THDM,
respectively, and $\theta_W$ is the \MSbar\ weak mixing angle.  The
coupling constants of the THDM are related to the corresponding
Standard Model ones, $\alpha_{\text{em}}^{\SM(5),\MSbar}(M_t)$ and
$\alpha_{\text{s}}^{\SM(5),\text{\MSbar}}(M_t)$, via the relation
\begin{align}
  \alpha_{\text{em}}^{\THDM}(M_t) &=
  \frac{\alpha_{\text{em}}^{\SM(5),\MSbar}(M_t)}{1 - \Delta\alpha_{\text{em}}(M_t)} \,,\\
  \alpha_{\text{s}}^{\THDM}(M_t) &=
  \frac{\alpha_{\text{s}}^{\SM(5),\text{\MSbar}}(M_t)}{1 - \Delta\alpha_{\text{s}}(M_t)} \,,
\end{align}
where the threshold corrections $\Delta\alpha_i(\mu_r)$ read
\begin{align}
  \Delta\alpha_{\text{em}}(\mu_r) &=
  \frac{\alpha_\text{em}}{2\pi} \left[
    - \frac{16}{9} \log{\frac{m_t}{\mu_r}}
    - \frac{4}{3} \sum_{i=1}^2 \log{\frac{m_{\tilde{\chi}^\pm_i}}{\mu_r}}
    - \frac{1}{3} \log{\frac{m_{H^\pm}}{\mu_r}}
  \right] \,,\\
  \Delta\alpha_{\text{s}}(\mu_r) &=
  \frac{\alpha_\text{s}}{2\pi} \left[
    - \frac{2}{3} \log{\frac{m_t}{\mu_r}}
    - 2 \log{\frac{m_{\tilde{g}}}{\mu_r}}
  \right] \,.
\end{align}
The terms involving the masses of the charginos and the gluino are
only present if these particles have not been integrated out at the
high-scale and are thus part of the low-energy effective theory.
As input, we use $\alpha_{\text{em}}^{\SM(5),\MSbar}(M_Z) = 1/127.940$
\cite{PDG}
and $\alpha_{\text{s}}^{\SM(5),\text{\MSbar}}(M_Z) = 0.1184$
\cite{Bethke:2009jm}, which are evolved to the scale $M_t$ using the
1-loop QED and 3-loop QCD $\beta$-functions in the Standard Model with
$5$ active quark flavours.

The \MSbar\ weak mixing angle $\theta_W$ in the THDM is determined
from the Fermi constant $G_F = 1.16638 \cdot 10^{-5}$ \cite{Agashe:2014kda} and $M_Z =
91.1876 \unit{GeV}$ \cite{Agashe:2014kda} using the iterative approach described in
\cite{Pierce:1996zz} taking into account the full 1-loop THDM corrections and
leading 2-loop Standard Model corrections to $\Delta\hat{\rho}$ and
$\Delta\hat{r}$ \cite{Fanchiotti:1992tu,Pierce:1996zz}.  The vertex and box contributions,
$\delta_{\text{VB}}$, from potential non-Standard Model particles are neglected here.

The \MSbar\ Yukawa couplings $y_i(M_t)$ of the THDM are determined
from the corresponding THDM \MSbar\ masses $m_i$ using the relations
\begin{align}
  y_i(M_t) =
  \begin{cases}
    m_i(M_t)/v_u(M_t) & \text{if $i$ is an up-type fermion} \,, \\
    m_i(M_t)/v_d(M_t) & \text{if $i$ is a down-type fermion} \,.
  \end{cases}
\end{align}
The top quark \MSbar\ mass in the THDM is calculated from the top pole
mass $M_t = 173.34 \unit{GeV}$ \cite{ATLAS:2014wva} using the full
1-loop self-energy plus 2-loop Standard Model QCD corrections,
\begin{align}
\begin{split}
  m_{t}(M_t) &= M_t + \re\Sigma_{t}^S(p^2=M_t^2,\mu_r=M_t) \\
  &\phantom{={}} + M_t \Big[ \re\Sigma_{t}^L(p^2=M_t^2,\mu_r=M_t) +
    \re\Sigma_{t}^R(p^2=M_t^2,\mu_r=M_t) \\
  &\phantom{={} + M_t \Big[}
    + \Delta m_t^{(1),\text{qcd}}(M_t) + \Delta m_t^{(2),\text{qcd}}(M_t) \Big]
  \,,
\end{split}
\end{align}
where $\Sigma_{t}^{S,L,R}$ denote the scalar, left- and right-handed
parts of the top self-energy in the \MSbar\ scheme
without the gluon contribution, and $\Delta
m_t^{(1),\text{qcd}}$ and $\Delta m_t^{(2),\text{qcd}}$ are 1- and
2-loop gluon corrections taken from Ref.~\cite{Fleischer:1998dw},
\begin{align}
  \Delta m_t^{(1),\text{qcd}}(\mu_r) &=
  -\frac{g_3^2}{12 \pi^2} \left[4 - 3 \log\left(\frac{m_t^2}{\mu_r^2}\right)\right],
  \label{eq:top-selfenergy-qcd-1L}\\
  \begin{split}
    \Delta m_t^{(2),\text{qcd}}(\mu_r) &= \left(\Delta
      m_t^{(1),\text{qcd}}\right)^2 - \frac{g_3^4}{4608 \pi^4}
    \Bigg[396 \log^2\frac{m_t^2}{\mu_r^2} - 2028 \log\frac{m_t^2}{\mu_r^2}
    - 48 \zeta(3) \\
    &\phantom{={}} + 2821 + 16 \pi^2 (1+\log 4)\Bigg] \,.
  \end{split}
  \label{eq:top-selfenergy-qcd-2L}
\end{align}
The bottom quark \MSbar\ mass in the THDM, $m_b(M_t)$, is obtained
from the \MSbar\ mass $m_b^{\SM(5)}(m_b) = 4.18 \unit{GeV}$ in the
Standard Model with $5$ active quark flavours by first evolving
$m_b^{\SM(5)}(m_b)$ to the scale $M_t$ using the 1-loop QED and 3-loop QCD RGE.
Afterwards, $m_b^{\SM(5)}(M_t)$ is converted to $m_b(M_t)$ as
\begin{align}
  m_b(M_t) &= \frac{m_b^{\SM(5)}(M_t)}{1 - \Delta m_b} \,, \\
  \Delta m_b &= \re\Sigma_{b}^S(p^2=m_b^2,\mu_r=M_t)/m_b \notag\\
  &\phantom{={}}+ \re\Sigma_{b}^L(p^2=m_b^2,\mu_r=M_t) +
\re\Sigma_{b}^R(p^2=m_b^2,\mu_r=M_t) \,,
  \label{eq:bottom-conversion}
\end{align}
where $\Sigma_{b}^{S,L,R}$ are the scalar, left- and right-handed
parts of the 1-loop bottom quark self-energy 
in the \MSbar\ scheme
in which
all Standard Model particles, except the bottom quark, the top quark and 
the W, Z and Higgs
bosons, are omitted.
Finally, the \MSbar\ mass of
the $\tau$ lepton in the THDM, $m_\tau(M_t)$, is calculated by first
identifying the $\tau$ pole mass, $M_\tau$, with the \MSbar\ mass in
the Standard Model with 5 active quark flavours at the scale $M_\tau$,
\begin{align}
  m_\tau^{\SM(5)}(M_\tau) &=  M_\tau \,.
\end{align}
In this identification, the 1-loop Standard Model electroweak
corrections to $m_\tau^{\SM(5)}(M_\tau)$ are neglected.
Afterwards, $m_\tau^{\SM(5)}(M_\tau)$ is evolved to $M_t$ using the
1-loop QED RGE and $m_\tau^{\SM(5)}(M_t)$ is converted to
$m_\tau(M_t)$ as
\begin{align}
\begin{split}
  m_\tau(M_t) &= m_\tau^{\SM(5)}(M_t)
  + \re\Sigma_{\tau}^S(p^2=m_\tau^2,\mu_r=M_t) \\
  &\phantom{={}}
  + m_\tau^{\SM(5)}(M_t) \Big[
    \re\Sigma_{\tau}^L(p^2=m_\tau^2,\mu_r=M_t)
    + \re\Sigma_{\tau}^R(p^2=m_\tau^2,\mu_r=M_t) \Big]\,,
\end{split}
\end{align}
where $\Sigma_{\tau}^{S,L,R}$ are the scalar, left- and right-handed
parts of the 1-loop $\tau$ self-energy in the \MSbar\ scheme
where all Standard Model
particles, except the $\tau$ lepton, the top quark and the W, Z and Higgs
bosons, are omitted.

The \MSbar\ vacuum expectation values $v_u(M_t)$ and $v_d(M_t)$ are
obtained from the running Z mass, $m_Z(M_t)$ and the \MSbar\ gauge
couplings via
\begin{align}
  v_d(M_t) &= \frac{\sqrt{2} m_Z(M_t)}{\sqrt{3/5 g_1^2(M_t) + g_2^2(M_t)} \cos\beta(M_t)} \,, \\
  v_u(M_t) &= \frac{\sqrt{2} m_Z(M_t)}{\sqrt{3/5 g_1^2(M_t) + g_2^2(M_t)} \sin\beta(M_t)} \,,
\end{align}
where the running Z mass is given by
\begin{align}
  m_Z^2(M_t) = M_Z^2 + \re\Sigma_{ZZ}^T(p^2=M_Z^2,\mu_r=M_t) ,
\end{align}
and $\Sigma_{ZZ}^T$ is the transverse part of the 1-loop Z self-energy
in the THDM including higgsinos and gauginos if present in the theory.

As shown above, the matching at the weak scale at the 1- and
2-loop level introduces a dependency of the gauge and Yukawa couplings
as well as the vacuum expectation values on the particle spectrum of
the THDM (possibly including higgsinos and gauginos).  These gauge and
Yukawa couplings enter the renormalisation group equations for all
model parameters, including the quartic couplings $\lambda_i$, which
are fixed by boundary conditions at the high scale, $M_S$.  For this
reason, an iteration between the matching of the $\lambda_i$ at $M_S$
and the matching to the Standard Model at $M_t$ must be performed
until a convergent solution to this boundary value problem has been
found.

If a consistent solution to this boundary value problem has been
found, the pole mass spectrum is calculated at the 1-loop level.  This
calculation follows a similar procedure as described in
Ref.~\cite{Pierce:1996zz} for the MSSM, adapted to the THDM case,
potentially including higgsinos and gauginos, if present in the theory.

\subsection{Vacuum (meta)stability}
\label{appendix:metastability}

Absolute stability of the electroweak vacuum is a strong
requirement. From the phenomenological point of view, it might be more reasonable
to demand metastability with a lifetime larger than the age of
the universe. Semiclassically, the tunnelling probability into the true vacuum during
a cosmic time $\tau$ (or more precisely, the tunnelling rate times $\tau$)
can be estimated as \cite{Coleman:1977py}
\be
p\sim\left(\frac{\tau}{R}\right)^4 e^{-S_{\rm bounce}}\,,
\ee
where $S_{\rm bounce}$ is the euclidean action
of the ``bounce'' instanton solution which interpolates between the false and the true
vacuum, and $R$ is the characteristic size of the bubble. Note that, at this level,
$R$ is undetermined for a classically scale invariant potential.

A more precise estimate in quantum theory was discussed e.g.~in Ref.~\cite{Isidori:2001bm}
for the case of the Standard Model. Following their analysis, for a single scalar
field with a $\phi^4$ potential (neglecting the Higgs mass term),
\be
{\cal L}=\frac{1}{2}(\partial_\mu\phi)^2-\frac{\lambda}{4}\phi^4\,,
\ee
the tunnelling probability for negative $\lambda$ can be estimated as
\be\label{eq:tunnelling}
p\approx\max_R\left(\frac{\tau}{R}\right)^4\exp\left(-\frac{8\pi^2}{3|\lambda(\frac{1}{R})|}+\Delta S\right)\,,
\ee
where $\lambda(\mu_r)$ is the running quartic coupling, and
$\Delta S$ are one-loop corrections from particles coupling to $\phi$. We require
$p\ll 1$ when $\tau$ is the age of the universe, $\tau=10^{10}$ yr. The tunnelling probability
is dominated by the largest value of $|\lambda|$, which,
for the Standard Model, leads to a condition that $\lambda$ be larger than about $-0.05$ during
its entire RG evolution up to $M_{\rm Planck}$  \cite{Isidori:2001bm}
(somewhat larger $|\lambda|$ being permissible at low scales).

In our case the model is somewhat more complicated as it involves several scalar degrees of freedom.
However, out of the four conditions for absolute stability Eqs.~(\ref{eq:stability1}-\ref{eq:stability4}), the first three turn
out always to be satisfied as a consequence of the supersymmetric boundary conditions on the quartics.
The remaining condition Eq.~\eqref{eq:stability4}
\be\nonumber
\tilde\lambda\equiv \lambda_3+\lambda_4+\sqrt{\lambda_1\lambda_2}>0
\ee
may be violated, which corresponds to one particular direction in field space becoming unstable.
To see this explicitly, we follow Ref.~\cite{Gunion:2002zf} and set
\be
a=H_1^\dag H_1\,,\qquad b=H_2^\dag H_2\,,\qquad c=\re H_1^\dag H_2\,,\qquad d=\im H_1^\dag H_2\,.
\ee
This allows us to write the quartic potential as the sum of three terms which are manifestly positive definite if
the stability conditions Eqs.~(\ref{eq:stability1}-\ref{eq:stability4}) are satisfied:
\be\label{eq:V4}
V_4=\frac{1}{2}\left(\sqrt{\lambda_1} a-\sqrt{\lambda_2}b\right)^2+(\lambda_3+\sqrt{\lambda_1\lambda_2})(ab-c^2-d^2)+\tilde\lambda (c^2+d^2)\,.
\ee
If however $\tilde\lambda$ is negative, then the potential is unbounded from below along the direction $a=\sqrt{\lambda_2/\lambda_1}b$, $ab=c^2+d^2$ with $c^2+d^2$ growing large.

To map this onto a one-dimensional problem, we choose a gauge and a field basis such that
\be
H_1=\left(\begin{array}{c} 0 \\ \frac{1}{\sqrt{2}}(\phi\,\cos\theta+\chi\,\sin\theta) e^{i\xi_1} \end{array}\right)\,,\qquad H_2=\left(\begin{array}{c} \frac{\rho}{\sqrt{2}}\, e^{i\xi_2} \\ \frac{1}{\sqrt{2}}(-\phi\,\sin\theta+\chi\,\cos\theta)e^{i\xi_3}\end{array}\right)\,,
\ee
where $\phi,\chi,\rho,$ and $\xi_i$ are real and $\theta$ is defined by
\be
\frac{1+\sin(2\theta)}{1-\sin(2\theta)}=\sqrt{\frac{\lambda_2}{\lambda_1}}\,.
\ee
Choosing $\rho=0$ and $\chi=\phi$ sets the first two terms in Eq.~\eqref{eq:V4} to zero. The remaining effective potential along the $\phi$ direction is
\be
V_{\rm eff}(\phi)=\frac{\tilde\lambda}{4}\cos^2(2\theta)\phi^4\,,
\ee
or equivalently
\be\label{eq:Veff}
V_{\rm eff}(\phi)=\frac{\lambda}{4}\phi^4\,,\qquad\text{where }\lambda=\frac{4\,\sqrt{\lambda_1\lambda_2}\,(\lambda_3+\lambda_4+\sqrt{\lambda_1\lambda_2})}{\lambda_1+\lambda_2+2\,\sqrt{\lambda_1\lambda_2}}\,.
\ee
The tunnelling rate will be dominated by bounces along this line in field space, so the problem is effectively one-dimensional. Using for $S_{\rm bounce}$ the RG-improved one-dimensional expression without further loop corrections,
\be
S_{\rm bounce}=\frac{8\pi^2}{3|\lambda(\mu_r)|}\,,
\ee
we obtain a reasonably accurate necessary condition for the longevity of the electroweak vacuum  from Eq.~\eqref{eq:tunnelling}. The condition is that at all scales $\mu_r$ between the electroweak scale and $M_S$ we should have the inequality
\be
\lambda(\mu_r)\gtrsim - \frac{2.82}{41.1+\log_{10}\frac{\mu_r}{\rm GeV}}\,,
\ee
with $\lambda$ defined in Eq.~\eqref{eq:Veff}. This lower bound on $\lambda$ varies between $-0.065$ at the electroweak scale and $-0.047$ at $\mu_r=M_{\rm Planck}$. It could probably be strengthened slightly by going beyond our simple one-dimensional approximation.



\begin{thebibliography}{99}

\bibitem{Green:1987mn}
  M.~B.~Green, J.~H.~Schwarz and E.~Witten,
  {\it Superstring Theory. Vol. 2: Loop Amplitudes, Anomalies And Phenomenology},
  Cambridge Univ.~Pr.~(1987).

\bibitem{Ibanez:2012zz}
  L.~E.~Ibanez and A.~M.~Uranga,
  {\it String theory and particle physics: An introduction to string phenomenology},
  Cambridge Univ.~Pr.~(2012).

\bibitem{Hall:2009nd}
  L.~J.~Hall and Y.~Nomura,
  JHEP {\bf 1003} (2010) 076
  [arXiv:0910.2235 [hep-ph]].

\bibitem{ArkaniHamed:2004fb}
N.~Arkani-Hamed and S.~Dimopoulos,
  JHEP
  {\bf 0506} (2005) 073
[arXiv:hep-th/0405159].

\bibitem{Giudice:2004tc}
G.~Giudice and A.~Romanino,
  Nucl.\ Phys.\ B {\bf 699} (2004) 65
[arXiv:hep-ph/0406088].

\bibitem{Antoniadis:2004dt}
  I.~Antoniadis and S.~Dimopoulos,
  Nucl.\ Phys.\ B {\bf 715} (2005) 120
  [hep-th/0411032].

\bibitem{Buchmuller:2015jna}
  W.~Buchmuller, M.~Dierigl, F.~Ruehle and J.~Schweizer,
  Phys.\ Lett.\ B {\bf 750} (2015) 615
  doi:10.1016/j.physletb.2015.09.069
  [arXiv:1507.06819 [hep-th]].

\bibitem{Giudice:2011cg}
  G.~F.~Giudice and A.~Strumia,
  Nucl.\ Phys.\ B {\bf 858} (2012) 63
  doi:10.1016/j.nuclphysb.2012.01.001
  [arXiv:1108.6077 [hep-ph]].

\bibitem{Bagnaschi:2014rsa}
  E.~Bagnaschi, G.~F.~Giudice, P.~Slavich and A.~Strumia,
  JHEP {\bf 1409} (2014) 092
  [arXiv:1407.4081 [hep-ph]].


\bibitem{Buttazzo:2013uya}
  D.~Buttazzo, G.~Degrassi, P.~P.~Giardino, G.~F.~Giudice, F.~Sala, A.~Salvio and A.~Strumia,
  JHEP {\bf 1312} (2013) 089
  doi:10.1007/JHEP12(2013)089
  [arXiv:1307.3536 [hep-ph]].

\bibitem{Gorbahn:2009pp}
  M.~Gorbahn, S.~Jager, U.~Nierste and S.~Trine,
  Phys.\ Rev.\ D {\bf 84} (2011) 034030
  doi:10.1103/PhysRevD.84.034030
  [arXiv:0901.2065 [hep-ph]].

\bibitem{Lee:2015uza}
  G.~Lee and C.~E.~M.~Wagner,
  Phys.\ Rev.\ D {\bf 92} (2015) 7,  075032
  doi:10.1103/PhysRevD.92.075032
  [arXiv:1508.00576 [hep-ph]].

\bibitem{Chakrabarty:2014aya}
  N.~Chakrabarty, U.~K.~Dey and B.~Mukhopadhyaya,
  JHEP {\bf 1412} (2014) 166
  [arXiv:1407.2145 [hep-ph]].

\bibitem{Das:2015mwa}
  D.~Das and I.~Saha,
  Phys.\ Rev.\ D {\bf 91} (2015) 9,  095024
  [arXiv:1503.02135 [hep-ph]].

\bibitem{Chowdhury:2015yja}
  D.~Chowdhury and O.~Eberhardt,
  arXiv:1503.08216 [hep-ph].

\bibitem{Ferreira:2015rha}
  P.~Ferreira, H.~E.~Haber and E.~Santos,
  Phys.\ Rev.\ D {\bf 92} (2015) 033003
  [arXiv:1505.04001 [hep-ph]].

\bibitem{Athron:2014yba}
  P.~Athron, J.-h.~Park, D.~St\"{o}ckinger and A.~Voigt,
  Comput.\ Phys.\ Commun.\  {\bf 190} (2015) 139
  [arXiv:1406.2319 [hep-ph]].

\bibitem{Amhis:2014hma}
  Y.~Amhis {\it et al.} [Heavy Flavor Averaging Group (HFAG) Collaboration],
  arXiv:1412.7515 [hep-ex].

\bibitem{Aad:2014vgg}
  G.~Aad {\it et al.} [ATLAS Collaboration],
  JHEP {\bf 1411} (2014) 056
  doi:10.1007/JHEP11(2014)056
  [arXiv:1409.6064 [hep-ex]].

\bibitem{CMS:2015mca}
  CMS Collaboration [CMS Collaboration],
  CMS-PAS-HIG-14-029.

\bibitem{ATLASHtautau13TeV}
  The ATLAS collaboration,
  ATLAS-CONF-2015-061.

\bibitem{Haber:1993an}
  H.~E.~Haber and R.~Hempfling,
  Phys.\ Rev.\ D {\bf 48} (1993) 4280
  [hep-ph/9307201].

\bibitem{Deshpande:1977rw}
  N.~G.~Deshpande and E.~Ma,
  Phys.\ Rev.\ D {\bf 18} (1978) 2574.


\bibitem{Staub:2008uz}
  F.~Staub,
  arXiv:0806.0538 [hep-ph].


\bibitem{Staub:2010jh}
  F.~Staub,
  Comput.\ Phys.\ Commun.\  {\bf 182} (2011) 808
  [arXiv:1002.0840 [hep-ph]].

\bibitem{Staub:2013tta}
  F.~Staub,
  Comput.\ Phys.\ Commun.\  {\bf 185} (2014) 1773
  [arXiv:1309.7223 [hep-ph]].

\bibitem{Lyonnet:2013dna}
  F.~Lyonnet, I.~Schienbein, F.~Staub and A.~Wingerter,
  Comput.\ Phys.\ Commun.\  {\bf 185} (2014) 1130
  doi:10.1016/j.cpc.2013.12.002
  [arXiv:1309.7030 [hep-ph]].

\bibitem{Lyonnet:2015jca}
  F.~Lyonnet,
  arXiv:1510.08841 [hep-ph].


\bibitem{Pierce:1996zz}
  D.~M.~Pierce, J.~A.~Bagger, K.~T.~Matchev and R.~j.~Zhang,
  Nucl.\ Phys.\ B {\bf 491} (1997) 3
  [hep-ph/9606211].

\bibitem{Fleischer:1998dw}
  J.~Fleischer, F.~Jegerlehner, O.~V.~Tarasov and O.~L.~Veretin,
  Nucl.\ Phys.\ B {\bf 539} (1999) 671
   [Nucl.\ Phys.\ B {\bf 571} (2000) 511]
  [hep-ph/9803493].


\bibitem{Coleman:1977py}
  S.~R.~Coleman,
  Phys.\ Rev.\ D {\bf 15} (1977) 2929
   [Phys.\ Rev.\ D {\bf 16} (1977) 1248].


\bibitem{Isidori:2001bm}
  G.~Isidori, G.~Ridolfi and A.~Strumia,
  Nucl.\ Phys.\ B {\bf 609} (2001) 387
  [hep-ph/0104016].

\bibitem{Gunion:2002zf}
  J.~F.~Gunion and H.~E.~Haber,
  Phys.\ Rev.\ D {\bf 67} (2003) 075019
  [hep-ph/0207010].

\bibitem{PDG}
  K.~A.~Olive {\it et al.} [Particle Data Group Collaboration],
  Chin.\ Phys.\ C {\bf 38} (2014) 090001.

\bibitem{Bethke:2009jm}
  S.~Bethke,
  Eur.\ Phys.\ J.\ C {\bf 64} (2009) 689
  [arXiv:0908.1135 [hep-ph]].

\bibitem{ATLAS:2014wva}
  [ATLAS and CDF and CMS and D0 Collaborations],
  arXiv:1403.4427 [hep-ex].

\bibitem{Agashe:2014kda}
  K.~A.~Olive {\it et al.} [Particle Data Group Collaboration],
  Chin.\ Phys.\ C {\bf 38} (2014) 090001.

\bibitem{Fanchiotti:1992tu}
  S.~Fanchiotti, B.~A.~Kniehl and A.~Sirlin,
  Phys.\ Rev.\ D {\bf 48} (1993) 307
  [hep-ph/9212285].

\end{thebibliography}
\end{document}